\documentclass[a4paper,onecolumn,11pt,accepted=2024-10-11]{quantumarticle}
\pdfoutput=1
\usepackage[utf8]{inputenc}
\usepackage[english]{babel}
\usepackage[T1]{fontenc}

\usepackage{tikz}
\usepackage{lipsum}

\usepackage{amssymb,amsmath,graphicx,color,dsfont, mathrsfs ,tikz-cd}
\usepackage{braket}
\usepackage{bm}
\usepackage{enumerate}
\usepackage{multirow}
\usepackage[caption=false]{subfig}
\usepackage[export]{adjustbox}
\usepackage{amsfonts}
\usepackage{amsthm}
\usepackage{adjustbox}
\usepackage{geometry}
\usepackage{array}
\usepackage{xfrac}
\usepackage{quantikz}
\usepackage{url}
\usepackage{xcolor}
\usepackage{rotating,booktabs}
\usepackage[verbose]{placeins}
\usepackage{ragged2e}
\usepackage[titletoc]{appendix}
\usepackage{svg}
\usepackage{slashed}
\usepackage{comment}

\usepackage[square,numbers,comma,sort&compress]{natbib}
\usepackage{hyperref}

\begin{document}

\title{Scattering wave packets of hadrons in gauge theories: Preparation on a quantum computer
}

\author{Zohreh Davoudi}
\email{davoudi@umd.edu}
\affiliation{Department of Physics, University of Maryland, College Park, MD 20742 USA}
\affiliation{Maryland Center for Fundamental Physics, University of Maryland, College Park, MD 20742, USA}
\affiliation{Joint Center for Quantum Information and Computer Science, National Institute of Standards and Technology (NIST) and University of Maryland, College Park, MD 20742 USA}
\affiliation{The NSF Institute for Robust Quantum Simulation, University of Maryland, College Park, Maryland 20742, USA}
\affiliation{National Quantum Laboratory (QLab), University of Maryland, College Park, MD 20742 USA}
\orcid{0000-0002-7288-2810}

\author{Chung-Chun Hsieh}
\email{cchsieh@umd.edu}
\thanks{corresponding author.}
\affiliation{Department of Physics, University of Maryland, College Park, MD 20742 USA}
\affiliation{Maryland Center for Fundamental Physics, University of Maryland, College Park, MD 20742, USA}
\orcid{0009-0004-8654-2034}

\author{Saurabh V. Kadam}
\email{ksaurabh@uw.edu}
\affiliation{
InQubator for Quantum Simulation (IQuS), Department of Physics, University of Washington, Seattle, WA 98195, USA}
\orcid{0000-0001-9218-1600}

\begin{abstract}
Quantum simulation holds promise of enabling a complete description of high-energy scattering processes rooted in gauge theories of the Standard Model. A first step in such simulations is preparation of interacting hadronic wave packets. To create the wave packets, one typically resorts to adiabatic evolution to bridge between wave packets in the free theory and those in the interacting theory, rendering the simulation resource intensive. In this work, we construct a wave-packet creation operator directly in the interacting theory to circumvent adiabatic evolution, taking advantage of resource-efficient schemes for ground-state preparation, such as variational quantum eigensolvers. By means of an ansatz for bound mesonic excitations in confining gauge theories, which is subsequently optimized using classical or quantum methods, we show that interacting mesonic wave packets can be created efficiently and accurately using digital quantum algorithms that we develop. Specifically, we obtain high-fidelity mesonic wave packets in the $Z_2$ and $U(1)$ lattice gauge theories coupled to fermionic matter in 1+1 dimensions. Our method is applicable to both perturbative and non-perturbative regimes of couplings. The wave-packet creation circuit for the case of the $Z_2$ lattice gauge theory is built and implemented on the Quantinuum \texttt{H1-1} trapped-ion quantum computer using 13 qubits and up to 308 entangling gates. The fidelities agree well with classical benchmark calculations after employing a simple symmetry-based noise-mitigation technique. This work serves as a step toward quantum computing scattering processes in quantum chromodynamics. 
\end{abstract}

\maketitle

\newpage
\tableofcontents

\section{Introduction}
\label{sec:1}
\noindent
Scattering processes have long served as a unique probe into the subatomic world, and continue to be the focus of modern research in nuclear and high-energy physics. They led to the discovery of fundamental constituents of matter, and enabled verification of the Standard Model of particle physics---a quantum mechanical and relativistic description of the strong and electroweak interactions in nature. Present-day and future particle colliders will continue to shed light on the structure of matter in form of hadrons and nuclei (e.g., at the Electron-Ion Collider~\cite{Deshpande:2005wd,Accardi:2012qut,AbdulKhalek:2021gbh,Achenbach:2023pba}). They will also probe exotic phases of matter (e.g., at the Relativistic Heavy-Ion Collider~\cite{csernai1994introduction,Bzdak:2019pkr,Achenbach:2023pba,Lovato:2022vgq}), and will search for new particles and interactions beyond the Standard Model (e.g., at the Large Hadron Collider~\cite{Narain:2022qud,Bruning:2012zz,Shiltsev:2019rfl}). A wealth of analytical and numerical methods have been developed over several decades to confront theoretical predictions based on gauge field theories of the Standard Model with experimental scattering data. These methods have reached unprecedented accuracy and precision in both perturbative and non-perturbative regimes of couplings~\cite{ParticleDataGroup:2022pth,CTEQ:1993hwr,Altarelli:1989ue,dokshitzer1991basics,FlavourLatticeAveragingGroupFLAG:2021npn}. A highlight of such progress is the use of lattice-gauge-theory (LGT) methods to conduct first-principles studies of hadrons from the underlying theory of the strong force, quantum chromodynamics (QCD)~\cite{FlavourLatticeAveragingGroupFLAG:2021npn,Davoudi:2022bnl,USQCD:2022mmc}.

The LGT method involves working with quantum fields placed on discrete and finite spacetime volumes with an imaginary time, so as to employ efficient Monte-Carlo techniques to compute observables. The Euclidean nature of computations, nonetheless, precludes direct access to scattering amplitudes. Indirect methods exist, starting from finite-volume formalisms developed in Refs.~\cite{Luscher:1986pf,Luscher:1990ux}, but they have been limited to low-energy and low inelasticity processes~\cite{Briceno:2017max,Hansen:2019nir,Davoudi:2020ngi}. High-energy scattering of hadrons and nuclei are substantially more complex due to the composite nature of the colliding particles and a plethora of asymptotic final-state particles that are often produced. Beyond asymptotic scattering amplitudes, the evolution of matter as a function of time elapsed after the collision holds the key to yet-not-fully understood mechanisms of fragmentation,  hadronization, and thermalization in particle colliders and in early universe~\cite{Andersson:1983ia,Webber:1999ui,Accardi:2009qv,Albino:2008gy,Begel:2022kwp,Berges:2020fwq}. Unfortunately, perturbative and non-perturbative tools, with the aid of classical computing, have had limited success in providing a full first-principles description of scattering processes to date.  
\begin{figure}[t!]
    \centering
    \includegraphics[width=0.95\textwidth]{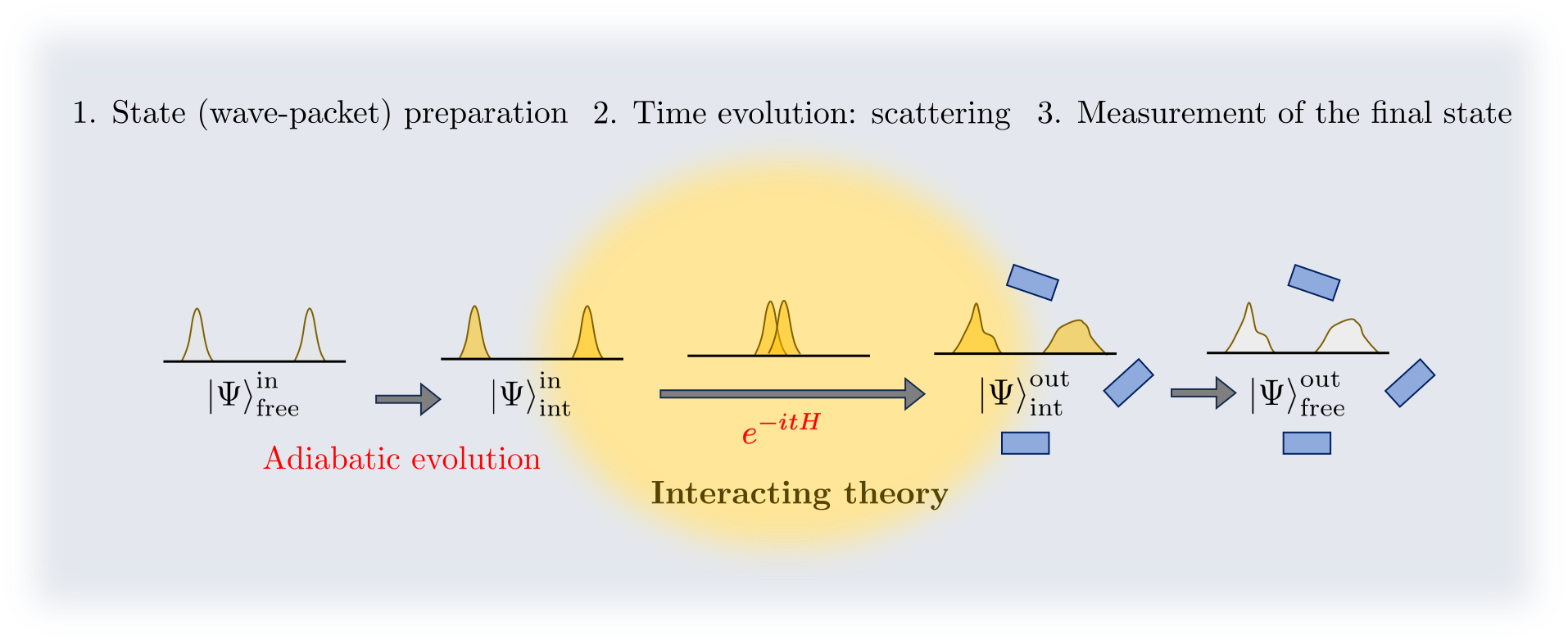}
    \caption{A schematic portrayal of JLP's protocol, a quantum algorithm for simulating scattering processes in the S-matrix formalism. The protocol consists of state preparation, time evolution under Hamiltonian $H$, and measurement. In the JLP protocol, adiabatic evolution is required to transform the incoming wave packets of the free theory, $\ket{\Psi}_{\rm free}^{\rm in}$, into those of an interacting theory, $\ket{\Psi}_{\rm int}^{\rm in}$. Similarly, outgoing wave packets of the interacting theory, $\ket{\Psi}_{\rm int}^{\rm out}$, are adiabatically turned into those of the free theory, $\ket{\Psi}_{\rm free}^{\rm out}$, before any measurement of the final state is performed. Alternatively, the state of the system can be measured at any post-collision stage in a quantum simulation.}
    \label{fig:JLP}
\end{figure}

Alternatively, one can resort to Hamiltonian simulation, whose real-time nature is deemed favorable for simulating scattering processes from first principles. On classical computers, Tensor-Network methods have proven efficient in simulating gauge theories in the Hamiltonian formalism~\cite{Banuls:2018jag,Banuls:2019rao,Meurice:2020pxc}, including for scattering processes in simple models~\cite{Pichler:2015yqa,Rigobello:2021fxw,Belyansky:2023rgh,Milsted:2020jmf,Van_Damme_2021}. However, the exponential growth of the Hilbert space as a function of system size, and the accumulation of unbounded entanglement in high-energy processes, are likely to make classical Hamiltonian simulation of gauge theories of the Standard Model infeasible. This motivates exploring the potential of simulating these theories on quantum hardware~\cite{Banuls:2019bmf,Halimeh:2023lid,Klco:2021lap,Bauer:2022hpo,Bauer:2023qgm,DiMeglio:2023nsa}. After mapping degrees of freedom of the system of interest to those of quantum platforms, time evolution can proceed in a digital or analog mode or a hybrid of both. The digital mode, which is the focus of this work, builds the unitary representing the Hamiltonian evolution out of a universal set of gates. The analog mode engineers a simulator Hamiltonian to mimic the target Hamiltonian, which is then evolved continuously. A hybrid mode combines features of both for more flexibility and efficiency. Digital~~\cite{Martinez:2016yna, Klco:2018kyo, Nguyen:2021hyk, Mueller:2022xbg, Chakraborty:2020uhf, deJong:2021wsd, Lamm:2019bik, Shaw:2020udc, Klco:2019evd, Atas:2021ext, Kan:2021xfc, Davoudi:2022xmb, Ciavarella:2021nmj, Atas:2022dqm, Paulson:2020zjd, Haase:2020kaj, Kane:2022ejm, Cohen:2021imf, Murairi:2022zdg, Davoudi:2022uzo, Farrell:2022wyt, Farrell:2022vyh, Charles:2023zbl, Mildenberger:2022jqr, Halimeh:2022pkw,Zhang:2023hzr,Kane:2023jdo, Hariprakash:2023tla},
analog~\cite{Zhou:2021kdl, Yang:2020yer, Mil:2019pbt, Banerjee:2012pg, Zohar:2012ts, Zohar:2012xf, Hauke:2013jga, Wiese:2013uua, Marcos:2013aya, Zohar:2015hwa, Yang:2016hjn, Bender:2018rdp, Davoudi:2019bhy, Luo:2019vmi, Notarnicola:2019wzb, Surace:2019dtp, Surace:2020ycc, Kasper:2020akk, Aidelsburger:2021mia, Andrade:2021pil, Surace:2023qwo},
and hybrid~\cite{Davoudi:2021ney, Gonzalez-Cuadra:2022hxt, Zache:2023cfj, Popov:2023xft, Meth:2023wzd}
schemes have been developed and implemented in recent years for increasingly more complex gauge theories. Most implementations, nonetheless, concern dynamics after a quantum quench~\cite{Mitra_2018}. A quench process involves preparing the simulation in a simple initial state and abruptly changing the Hamiltonian to the Hamiltonian of interest in order to create non-equilibrium conditions. In order to simulate scattering processes, however, one needs to initialize the quantum simulation in more complex states, such as particle wave packets.

Jordan, Lee, and Preskill (JLP) pioneered studies of scattering in quantum field theories on a quantum computer~\cite{Jordan:2012xnu, Jordan:2011ci}. Within an interacting scalar field theory, they laid down a systematic procedure for state preparation, time evolution, and measurement of wave packets. In particular, they considered the S-matrix formalism where the fields only interact in a scattering region, and are asymptotically free in the far past and the far future. Figure~\ref{fig:JLP} summarizes the JLP protocol. Here, the interacting wave packets are obtained by evolving the free wave packets with a time-dependent Hamiltonian where the interactions are turned on adiabatically. Theoretically, the success of this adiabatic interpolation depends on relevant energy gaps~\cite{Farhi:2000ikn}, hence difficulties are expected when the adiabatic path crosses phase transitions. It also requires extra care to compensate for the broadening of the wave packet during the preparation time~\cite{Jordan:2012xnu}. In practice, the protocol becomes infeasible as the required adiabatic evolution time often exceeds the decoherence time of the present quantum hardware. Under the adiabatic framework, it is possible to achieve better resource scalings, e.g., by using more natural bases for the Hilbert space in certain limits, such as low particle numbers~\cite{Barata:2020jtq}, or by utilizing efficient ways to traverse the adiabatic path in the Hamiltonian's parameter space~\cite{Cohen:2023dll}. Still, state preparation remains a bottleneck for general Hamiltonian simulation of scattering processes, particularly when the scattering particles are not excitations of the fundamental fields, and are rather composite (bound) states, as is the case in confining gauge theories.

Various proposals have been put forth in recent years to eliminate the need for adiabatic state preparation in scattering protocols. For example, in 1+1-dimensional systems, one may extract the elastic-scattering phase shift using real-time evolution in the early and intermediate stages of the collision using measured time delay of the wave packets~\cite{Gustafson:2021imb}. It is also possible to indirectly extract information about scattering by resorting to the relations between finite-volume spectrum of the interacting theory and scattering phases shifts~\cite{Luscher:1986pf,Luscher:1990ux,Luu:2010hw}. Here, quantum computers replace classical computers to compute the spectra, e.g., using hybrid classical-quantum schemes such as variational methods~\cite{Peruzzo:2013bzg,McClean:2015vup,Tilly:2021jem}, see e.g., Ref.~\cite{Sharma:2023bqu}. As already mentioned, this limits the scope of such studies to asymptotic amplitudes and low energies, and does not benefit from the full power of quantum computers. Alternatively, the full state preparation in the JLP protocol can be implemented via a variational quantum algorithm~\cite{Liu:2021otn}, assuming that suitable variational ansatze can be found for interacting excitations, which may in general be non-trivial. In another scheme, one may calculate scattering amplitudes directly from $n$-point correlation functions using the Lehmann–Symanzik–Zimmermann reduction formalism~\cite{Li:2023kex,Briceno:2023xcm,Briceno:2020rar}, as well as particle decay rates with the use of optical theorem~\cite{Ciavarella:2020vqm}, circumventing the need for wave-packet preparation. Nonetheless, creating wave-packet collisions that mimic actual scattering experiments allows probing the real-time dynamics of scattering, including entanglement generation and phases of matter, which are not directly accessible in asymptotic scattering amplitudes. Another alternative to circumvent adiabatic wave-packet preparation is to construct wave-packet operators directly in the interacting theory~\cite{Turco:2023rmx}, taking advantage of the Haag-Ruelle scattering theory~\cite{Haag:1958vt,Ruelle1962} developed within axiomatic framework of quantum field theory~\cite{Duncan:2012aja}, see also Ref.~\cite{Kreshchuk:2023btr} for a similar approach. Nonetheless, extensions of these frameworks to theories with massless excitations, including gauge theories, are yet to be developed. Recently, an algorithm was developed and implemented on a quantum hardware which prepares wave packets of (anti)fermions in a fermionic lattice field theory in 1+1 dimensions (D)~\cite{Chai:2023qpq}. Fermionic excitations are generated from the interacting vacuum, which is itself prepared by a variational algorithm. Still, a practical algorithm for preparation of bound-state wave packets in gauge theories is desired.
\begin{figure}[t!]
    \centering
    \includegraphics[width=1\textwidth]{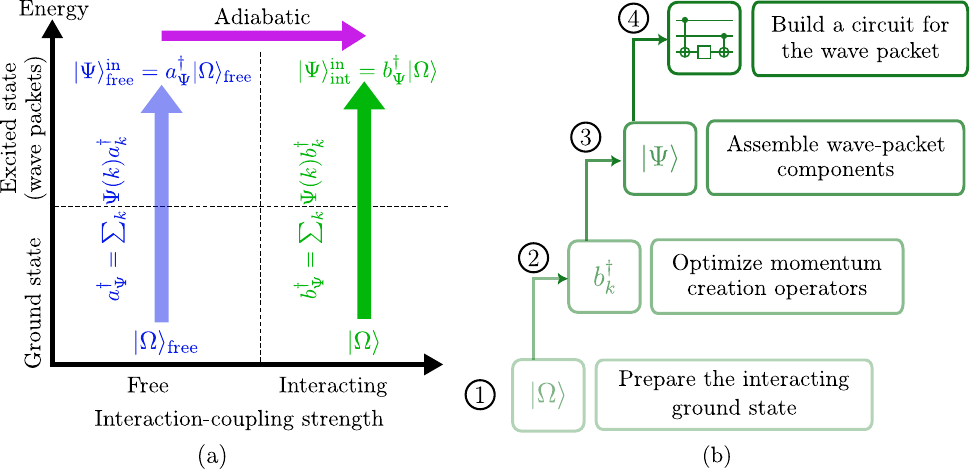}
    \caption{(a) A schematic comparison of the approach of this work and adiabatic state preparation. The blue and magenta paths represent the wave-packet creation in the JLP formalism. In order to arrive at the interacting wave packet, the adiabatic evolution is applied to the free wave-packet state created by $a_\Psi^\dagger$, which acts on the free ground state, $\ket{\Omega}_{\rm free}$. The algorithm proposed in this work, the green path, directly builds the interacting wave-packet creation operator $b_\Psi^\dagger$ that acts on the interacting ground state, $\ket{\Omega}$. $\ket{\Psi}_{\rm free (int)}^{\rm in}$ denotes the incoming wave packet in the free (interacting) theory. $\Psi(k)$ denotes the wave-packet profile and $k$ is the momentum, see Sec.~\ref{sec:2} for details. (b) A summary of various steps of the algorithm of this work, as detailed in Sec.~\ref{sec:3-1}.}
    \label{fig:schematic}
\end{figure}

Capitalizing on the idea of creating wave packets directly in the interacting theory, and inspired by a construction using tensor-network methods to study scattering in the lattice Schwinger Model~\cite{Rigobello:2021fxw}, we propose an efficient hybrid classical-quantum algorithm that creates wave packets of bound excitations in confining gauge theories in 1+1 D. This method, therefore, requires no adiabatic interpolation. Figure~\ref{fig:schematic} provides a birds-eye view of our state-preparation algorithm compared to the JLP algorithm. Specifically, we use an efficient variational ansatz for the interacting ground state and verify its efficiency in the case of a $Z_2$ LGT with fermionic matter in 1+1 D. We further optimize a physically motivated ansatz for the meson creation operator out of such an interacting vacuum, which exhibits high fidelity in both perturbative and non-perturbative regimes of the couplings, as verified in the case of the $Z_2$ and $U(1)$ LGTs in 1+1 D. Our algorithm demands only a number of entangling gates that is polynomial in the system size, and uses a single ancilla qubit, benefiting from recently developed algorithms based on singular-value decomposition of operators~\cite{Davoudi:2022xmb}. The quantum circuit that prepares a single wave packet is constructed for the case of the $Z_2$ LGT. For 6 fermionic sites (12+1 qubits), this circuit is executed on Quantinuum's hardware, System Model \texttt{H1}, a quantum computer based on trapped-ion technology. Both the algorithmic and experimental fidelities are analyzed, and various sources of errors are discussed. We further discuss observables that can be measured efficiently in experiment to verify the accuracy of the generated wave packet. While (controlled) approximations are made to achieve shallower circuits, and a rather simple noise mitigation is applied based on symmetry considerations, high fidelities are still achieved in this small demonstration. Hence, quantum simulation of hadron-hadron collisions in lower-dimensional gauge theories may be within the reach of the current generation of quantum hardware.

This paper is organized as follows. In Sec.~\ref{sec:2-1}, we introduce the 1+1-dimensional LGTs coupled to staggered fermions. We then specify an interacting creation-operator ansatz in such theories in Sec.~\ref{sec:2-2}, and demonstrate its validity through a numerical study in the case of $Z_2$ LGT in Sec.~\ref{sec:2-3}. The state-preparation algorithm and the circuit design are detailed in Sec.~\ref{sec:3-1}, with a focus on the case of $Z_2$ LGT given its lower simulation cost. Section~\ref{sec:3-2} includes our results on the creation of wave packets with the use of both numerical simulators and a quantum computer. We end in Sec.~\ref{sec:summary} with a summary and outlook. A number of appendices are provided to provide further details on the ansatz validity, circuit performance, and quantum-emulator comparisons. All data associated with numerical optimizations and circuit implementations are provided in Supplemental Material~\footnote{The supplemental material can also be found at \url{https://bit.ly/2402-00840}.}.

\section{An ansatz for mesonic wave packets in gauge theories in 1+1 D
}
\label{sec:2}
\noindent
The goal of this work is to demonstrate a suitable wave-packet preparation method for gauge theories that exhibit confined excitations. To keep the presentation compact, we focus on the case of Abelian LGTs in 1+1 D, and further specialize our study to the $Z_2$ and $U(1)$ LGTs coupled to one flavor of staggered fermions. We will later comment on the modifications required in the construction of the ansatz to make it suitable for non-Abelian groups such as in $SU(2)$ and $SU(3)$ LGTs. This section briefly introduces the LGTs of this work, presents details of our mesonic creation-operator ansatz, and demonstrates the validity of the proposed operator upon numerical optimizations in small systems. This ansatz will form the basis of our quantum-circuit analysis and implementation in the next section.

\subsection{Models: $Z_2$ and $U(1)$ LGTs coupled to fermions in 1+1 D}
\label{sec:2-1}
The Hamiltonian of an Abelian LGT coupled to one flavor of staggered fermions in 1+1 D can be written in the generic form:
\begin{equation}
    H = \frac{1}{2} \sum_{n \in \Gamma}{\left( \xi_{n}^\dagger U_n \xi_{n+a} + \text{H.c.}\right)} + am_f\sum_{n \in \Gamma}{(-1)^{n/a}  \xi_{n}^\dagger\xi_n }+a\epsilon \sum_{n \in \Gamma}f(E_n).
    \label{eq:Abelian-H}
\end{equation}
Here, $\Gamma = \big \{0,a,\cdots,(N-1)a \big \}$ is the set of lattice-site coordinates. $a$ is the lattice spacing and $N$ denotes the number of staggered sites (and is hence even). $\xi^\dagger_n$ ($\xi_n$) stands for the fermionic creation (annihilation) operator at site $n$. In the staggered formulation, first developed in Refs.~\cite{Kogut:1974ag, Banks:1975gq}, fermions (antifermions) live on the even (odd) sites of the lattice while the links host the gauge bosons. $U_n$ and $E_n$ are non-commuting conjugate operators representing the gauge-link and the electric-field operators on the link emanating from site $n$. $m_f \geq 0 $ is the fermion mass and $\epsilon$ is the strength of the electric-field Hamiltonian, expressed with the function $f(E_n)$ for generality. For example, in the case of the $Z_2$ LGT, $f(E_n)=E_n$ while in the $U(1)$ case, $f(E_n)=E_n^2$. For the rest of this paper, we set $a=1$. The continuum limit is, therefore, realized in the limit of $m_f,\epsilon \to 0$.

\begin{figure}[t!]
    \centering
    \includegraphics[width=0.95\textwidth]{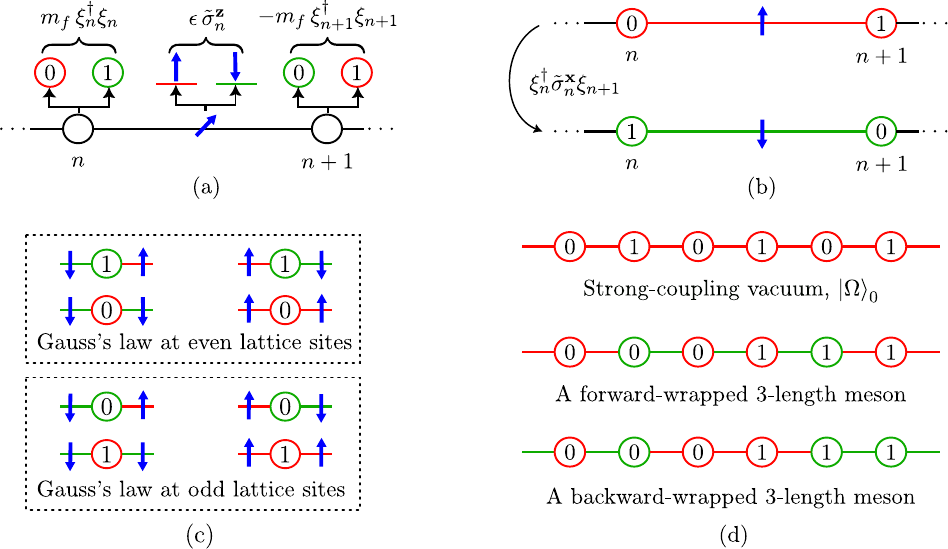}
    \caption{The degrees of freedom in the $Z_2$ LGT are shown in (a). A lattice site $n$ (taken to be even in this figure) and its neighboring lattice site $n+1$ are denoted by circles, and the gauge link connecting them is denoted by a line. A fermion (antifermion) at $n$ $(n+1)$ is represented by an occupied (unoccupied) lattice site, which is shown here by $1$ $(0)$ in a green circle, and the absence of a fermion (antifermion) at $n$ $(n+1)$ is shown by $0$ $(1)$ in a red circle. The corresponding fermion (antifermion) mass term in Eq.~\eqref{eq:Abelian-H} is written above the lattice site $n$ $(n+1)$. The spin-$\frac{1}{2}$ hardcore boson residing on the link is indicated by a blue arrow along with its electric-field Hamiltonian term given in Eq.~\eqref{eq:Abelian-H}.
    For $\epsilon<0$ which is used in the simulations of this work and assumed in the figure, the down spin (green link) has higher electric-field energy than the up spin (red link).
    The action of the operator $\xi^\dagger_{n}\tilde{\sigma}^{\textbf{x}}_{n}\xi_{n+1}$, which is a part of the fermion hopping Hamiltonian in Eq.~\eqref{eq:Abelian-H} for a $Z_2$ LGT, is depicted in (b). The upper (lower) dotted box in (c) shows configurations that satisfy the Gauss's law given in Eq.~\eqref{eq:gauss} for the case of the $Z_2$ LGT at an even (odd) lattice site. Three different states for a $Z_2$ LGT with PBCs and 6 lattice sites are shown in (d). The top picture represents the strong-coupling vacuum state, $\ket{\Omega}_0$, while the middle (bottom) picture depicts a forward-wrapped (backward-wrapped) 3-length bare-meson state created by the action of bare mesonic operators, defined in Sec.~\ref{sec:2-2}, on $\ket{\Omega}_0$.
    }
    \label{fig:Z2-schematic}
\end{figure}

Similarly, the specific form of the gauge-link and electric-field operators and their action on their respective local bosonic Hilbert space depend on the gauge group. In the case of the $Z_2$ LGT, the local Hilbert space in the electric-field basis is spanned by $\ket{s}$ with $s=\uparrow,\downarrow$, the two spin projections of a spin-$\frac{1}{2}$ hardcore boson along the $z$ axis, with $U=\ket{\uparrow}\bra{\downarrow}+\ket{\downarrow}\bra{\uparrow}\equiv\tilde{\sigma}^{\mathbf{x}}$ and $E=\ket{\uparrow}\bra{\uparrow}-\ket{\downarrow}\bra{\downarrow}\equiv\tilde{\sigma}^{\mathbf{z}}$. For the $U(1)$ LGT, the local Hilbert space in the electric field basis is the infinite-dimensional Hilbert space of a quantum rotor $\ket{\ell}$ with $\ell \in \mathbb{Z}$, with $U=\sum_\ell \ket{\ell+1}\bra{\ell}$ and $E=\sum_\ell \ell \ket{\ell}\bra{\ell}$. For practical purposes, the $\ell$ quantum number is cut off at a finite value $\Lambda > 0$, i.e., $-\Lambda \leq \ell \leq \Lambda$, up to an uncertainty that can be systematically controlled. The degrees of freedom and the action of the Hamiltonian terms in Eq.~\eqref{eq:Abelian-H} are illustrated in Figs.~\ref{fig:Z2-schematic} (a) and (b) for the case of the $Z_2$ LGT.

Only a portion of the Hilbert space spanned by the fermionic and bosonic basis states is physically relevant. This is because physical states of the theory must satisfy local Gauss's laws,
\begin{align}
    &G_n\ket{\psi_{\text{phys}}} = g \ket{\psi_{\text{phys}}}~\forall n,\label{eq:gauss}
\end{align}
with a specific value of the eigenvalue $g$. For the $Z_2$ LGT, $G_n = E_nE_{n-1}e^{-i\pi(\xi_n^\dagger\xi_n-\frac{1-(-1)^n}{2})}$ with $g=1$, while for the $U(1)$ LGT, $G_n = E_n-E_{n-1}+\xi_n^\dagger\xi_n-\frac{1-(-1)^n}{2}$ with $g=0$. The Gauss's law satisfying configurations are shown for the case of the $Z_2$ LGT in Fig.~\ref{fig:Z2-schematic}(c).

We will later build the interacting vacuum out of the strong-coupling vacuum, $\ket{\Omega}_0$, which is the ground state of Eq.~\eqref{eq:Abelian-H} in the limit of $\epsilon \gg 1$.
The fermion hopping is suppressed in the strong-coupling limit, and the lowest-energy configuration corresponds to the lowest-energy configuration of the electric field on all links with no fermion or antifermion present. For the $Z_2$ LGT, this implies all link spins pointing up (down) in the $z$ basis for $\epsilon < 0$ ($\epsilon > 0$), while for the $U(1)$ LGT, this implies that electric-field eigenvalues $\ell_n$ are zero at all links. Furthermore, the fermionic configuration in the strong-coupling vacuum is dictated by the sign of mass, taken to be positive in this work. The lowest-energy configuration corresponds to the eigenvalue of the $\xi^\dagger_n \xi_n$ operator being 0 (1) for no fermion (no antifermion) at even (odd) sites. 

We consider periodic boundary conditions (PBCs) throughout this paper, i.e., we impose $\xi_N=\xi_0$. PBCs ensure a parity symmetry on states and allows for specifying well-defined momentum quantum numbers, which is a key in constructing wave packets localized in momentum space~\footnote{For large lattice sizes, open boundary conditions can also be applied, with small modifications in defining momentum eigenstates.}. Nonetheless, when PBCs are imposed, one is restricted to retain gauge-field degrees of freedom in the simulation, i.e., Gauss's laws are not sufficient to eliminate the electric-field configuration throughout the lattice. Furthermore, PBCs in the $U(1)$ LGT imply that the total number of fermions must be equal to the total number of antifermions in the lattice for all states, since each fermion (antifermion) lowers (raises) the electric field strength on the gauge link emanating from its lattice site by one unit as seen from the Gauss's law in Eq.~\eqref{eq:gauss}. The strong-coupling vacuum satisfies this condition as it has zero fermions and antifermions.
Since the Hamiltonian in Eq.~\eqref{eq:Abelian-H} commutes with the total fermion-number operator $Q=\sum_{n=0}^{N-1} \xi^\dagger_n\xi_n$, all states that satisfy PBCs must have the same $Q$ eigenvalue as that of the strong-coupling vacuum, i.e., $N/2$. On the other hand, the states in the $Z_2$ LGT that satisfy PBCs have equal total number of fermions and antifermions modulo two. Throughout this paper, we restrict to the subspace of states that have $Q=N/2$ for both theories.

Finally, as will be introduced shortly, our ansatz is comprised of operators that, if acted on the strong-coupling vacuum, create `bare' mesons. These are fermion-antifermion excitations separated by a number of gauge links (a flux), specifying the meson `length'. With PBCs, there are always two such connecting fluxes in the `forward' (increasing index $n$) and `backward' (decreasing index $n$) directions. Examples of such mesons, along with the strong-coupling vacuum, in the $Z_2$ LGT are shown in Fig.~\ref{fig:Z2-schematic}(d). Clearly these states all satisfy the Gauss's law.

Last but not least, the Hamiltonian in Eq.~\eqref{eq:Abelian-H} is mapped to a qubit representation as follows. First, to properly treat the statistics of fermionic fields, we choose a Jordan-Wigner transform for fermions:
\begin{subequations}
\label{eq:jw}
\begin{align}
        \xi_n^\dagger = \left(\displaystyle \prod_{m=0}^{n-1}{\sigma^{\textbf{z}}_m}\right) \sigma_n^{-}, \label{eq:jwa}\\
        \xi_n = \left(\displaystyle \prod_{m=0}^{n-1}{\sigma^{\textbf{z}}_m}\right) \sigma_n^+.
    \label{eq:jwb}
\end{align}
\end{subequations}
With the PBCs, the fermion hopping term between lattice sites $N-1$ and 0 can be re-expressed as $ \sigma^-_{0}\sigma^+_{n-1} (-1)^{Q+1}+ \text{H.c.}$.
Here, the factor $(-1)^{Q+1}$ is obtained by pulling a string of ${\sigma^{\textbf{z}}}$ Paulis that span over the entire lattice to the right.
Furthermore, this factor evaluates to $(-1)^{N/2+1}$ following the discussion on PBCs given above.
Second, the local Hilbert space of each gauge link is mapped to a number of qubit registers depending on the gauge group under study and the bases chosen to express these degrees of freedom. The case of the $Z_2$ LGT is trivial, with one single qubit at each site representing the gauge boson. In the $U(1)$ LGT, one can use e.g., a binary encoding with $\lceil\log(2\Lambda+1)\rceil$ qubits at each link to represent the gauge degrees of freedom in the electric-field basis~\cite{Shaw:2020udc}.

\subsection{Ansatz for interacting wave-packet creation operator}
\label{sec:2-2}
Eigenstates of LGTs with PBCs, as discussed in the previous section, can be labeled by momentum quantum numbers. Let $\ket{k}$ be the lowest-energy state in each $k$-momentum sector with no overlap to the interacting vacuum. That is, this state is obtained by applying the interacting creation operator $b_k^\dagger$ to the interacting vacuum $\ket{\Omega}$:
\begin{equation}
    \ket{k} = b_k^\dagger\ket{\Omega},
    \label{eq:bk1}
\end{equation}
with a normalization chosen such that $\braket{\Omega|\Omega}=\braket{k|k}=1$. Then, a wave-packet state $\ket{\Psi}$ in momentum space is just the collection of states $\ket{k}$ smeared by the wave-packet profile $\Psi(k)$:
\begin{equation}
    \ket{\Psi} = \sum_k {\Psi(k) \ket{k}}. \label{eq:bk2}
\end{equation}
Here, we take the wave-packet profile to exhibit a Gaussian in momentum space with width $\sigma$ centered around $k_0$, with the corresponding Gaussian profile in position space centered around $\mu$,
\begin{equation}
    \Psi(k)=\mathcal{N}_\Psi \, \text{exp}\left(-ik\mu\right) \text{exp}\left(-\frac{(k-k_0)^2}{4{\sigma}^2}\right),
    \label{eq:Psi-def}
\end{equation}
with $\mu$, $\sigma$, and $k_0$ being real parameters. The normalization constant $\mathcal{N}_\psi$ is chosen such that $\braket{\Psi|\Psi}=1$.  One can also define the wave-packet creation operator $b_\Psi^\dagger$ in the same manner:
\begin{equation}
    b_{\Psi}^{\dagger} = \sum_k {\Psi(k) b_{k}^{\dagger}}.
    \label{eq:bk3}
\end{equation}
The procedure to generate $\ket{\Omega}$ will be described in Sec.~\ref{sec:3-1-1} for the example of the $Z_2$ LGT. The remainder of this section will be dedicated to the construction of $b_{k}^{\dagger}$ in each momentum sector, which eventually are assembled into the full wave-packet operator $b_{\Psi}^{\dagger}$.

The core of the interacting $b_{k}^{\dagger}$ is an ansatz proposed in Ref.~\cite{Rigobello:2021fxw}, in which wave packets of mesons in the lattice Schwinger model were built using tensor networks. This ansatz worked well only in the weak-coupling ($\epsilon \ll 1$) regime of the Schwinger model for system sizes studied in that work. As we will see, by adding a layer of classical or quantum optimization, this ansatz can be made suitable for both perturbative and non-perturbative regimes. To proceed, recall that in a confined gauge theory, $ b_{k}^{\dagger} $ should only create gauge-invariant mesonic bound states. One can model $ b_{k}^{\dagger} $ with a combination of mesonic creation operators consisting of fermion-antifermion creation-operator pairs:
\begin{align}
&b_{k}^{\dagger} = \sum_{p,q \in \widetilde{\Gamma}} \delta_{k, p+q} \eta(p, q) \mathcal{B}(p,q),
\label{eq:ansatz1}
\end{align}
with
\begin{align}
\mathcal{B}(p,q)=\sum_{m,n \in \Gamma} \mathscr{C}(p,m)\mathscr{D}(q,n) \mathcal{M}_{m,n}.
\label{eq:B-def}
\end{align}
Here, the momentum sums run over the Brillouin zone of the staggered lattice, $\widetilde{\Gamma} = \frac{2\pi}{N} \big\{-\frac{N}{2},-\frac{N}{2}+1, \cdots, \frac{N}{2}-1 \big\}\cap [-\frac{\pi}{2}, \frac{\pi}{2})$. 
$\eta(p,q)$ is a suitable function of constituent momenta $(p,q)$ and characterizes the ansatz, as will be discussed shortly.
Furthermore,
\begin{subequations}
\begin{align} 
&\mathscr{C}(p,m)=\sqrt{\frac{m_f+\omega_p}{2\pi\omega_p}} {e^{ipm}}\left(\mathcal{P}_{m0}+v_p\mathcal{P}_{m1}\right),
\\
&\mathscr{D}(q,n)=\sqrt{\frac{m_f+\omega_q}{2\pi\omega_q}} {e^{iqn}}\left(-v_q\mathcal{P}_{n0}+\mathcal{P}_{n1}\right).
\end{align}
\end{subequations}
Here, $\omega_k=\sqrt{m_f^2+\sin^2(k)}$, $v_k=\frac{\sin(k)}{m_f+\omega_k}$, and $\mathcal{P}_{n0(1)}=\frac
{1+(-1)^{n+0(1)}}{2}$ is the projection operator to the even (odd) staggered sites.

$\mathcal{M}_{m,n}$ is a gauge-invariant operator that, for the simple case of $m=n$, reduces to the fermion number operator $\xi^\dagger_m\xi_m$.
However, when $m\neq n$, $\mathcal{M}_{m,n}$ is given by a non-local operator, which we refer to as a bare-meson creation operator, that is composed of $\xi^\dagger_m$, $\xi_n$, and a string of $U$ or $U^\dagger$ operators connecting them.
Consider first the case $m<n$.
For the periodic lattice, there can be two ways of constructing such an operator: $\xi^\dagger_m (\prod_{l=m}^{n-1}U_l)\xi_n$ or $\xi^\dagger_m (\prod_{l=m-1}^{0}U_l^\dagger)(\prod_{l=N-1}^{n}U_l^\dagger)\,\xi_n$.
We call the former a forward-wrapped $(n-m)$-length meson creation operator and the latter a backward-wrapped $(N-n+m)$-length meson creation operator.
Similarly, when $m>n$, the forward-wrapped (backward-wrapped) meson creation operators are given by the same expression after replacing $U_l\to U^\dagger_l$ ($U^\dagger_l \to U_l$).
We require the ansatz to only build the shorter meson depending on $m$ and $n$.
If the meson length is equal for both forward and backward wrapping, i.e., $|m-n|=N/2$, each operator is added with a coefficient $\frac{1}{\sqrt{2}}$ such that each meson is created with a probability of $\frac{1}{2}$.
 
The form in Eq.~\eqref{eq:ansatz1} is motivated by the fact that for staggered free fermions (i.e., $U_n=1$ at all links), the expression in the square bracket in Eq.~\eqref{eq:ansatz1} reduces to $c_p^{\dagger} d_q^{\dagger}$, where $c_p^\dagger=\sum_m \mathscr{C}(p,m)\xi_m^\dagger$ and $d_q^\dagger=\sum_n \mathscr{D}(q,n)\xi_n$ diagonalize the free-fermion Hamiltonian in momentum space~\cite{Rigobello:2021fxw}.
Note that by explicitly constructing the global $U(1)$ charge operator, one finds that the $c_k$-type excitations are the antiparticles of the $d_k$-type excitations (have an opposite U(1) charge). Similarly, by constructing  the four-momentum operator, $P^\mu=(H,\bm{P})$, and demanding zero energy and momentum for such a state, it can be seen that $c_k$ and $d_k$ annihilate the vacuum~\cite{Rigobello:2021fxw}. 

\begin{figure}[t!]
    \centering
    \includegraphics[width=\textwidth]{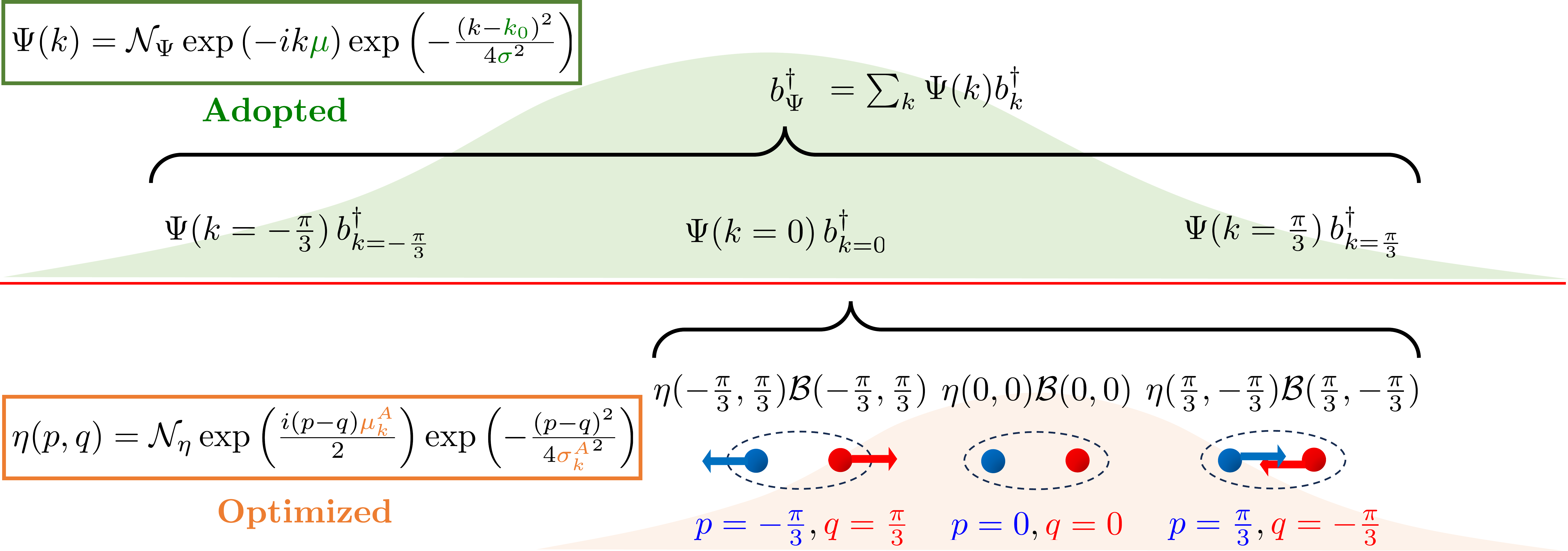}
    \caption{The schematic diagram of the double smearing of the mesonic wave packet. $b_\Psi^\dagger$ is assembled from $b_k^\dagger$ weighted with $\Psi(k)$, the manually-adopted wave-packet profile. Each $b_k^\dagger$ is built upon optimized mesonic ansatz $\eta(p,q)$. 
    }
    \label{fig:double_smearing}
\end{figure}

The fermion-antifermion pairs, or bare mesons, are distributed in momentum space following the ansatz function $\eta(p,q)$. Similar to  Ref.~\cite{Rigobello:2021fxw}, we consider a Gaussian distribution in relative momentum $p-q$:
\begin{equation}
\eta(p,q) = \mathcal{N}_\eta \, \text{exp}\left(\frac{i\mu_k^A(p-q)}{2}\right) \text{exp}\left(-\frac{(p-q)^2}{4{\sigma_k^A}^2}\right).
\label{eq:ansatz3}
\end{equation}
Here, $\mu_k^A$ and $\sigma_k^A$ are real parameters, superscript A denotes the ansatz, and $\mathcal{N}_\eta$ is the normalization factor. The Gaussian distribution ensures that a fermion and an antifermion with a large relative momentum are penalized. This is reasonable, as otherwise the constituents will eventually move far away from each other and would not form a bound excitation. $\mu_k^A$ controls the average separation of the fermion and antifermion in position space. Finally, because of the Kronecker delta in momenta in Eq.~\eqref{eq:ansatz1}, $p+q$ is forced to match the total momentum of the meson excitation, $k$.

In Ref.~\cite{Rigobello:2021fxw}, $b_k^\dagger\ket{\Omega}$ describes the momentum eigenstate $\ket{k}$ with $\sigma_k^A$ and $\mu_k^A$ manually tuned for each $k$. In this work, optimization on $\left(\sigma_k^A,\mu_k^A\right)$ is explicitly performed by searching for the lowest-energy state with $b_k^\dagger$ excitations in each $k$ sector. For small systems, $b_k^\dagger$ with the optimized parameters is benchmarked against exact-diagonalization results to ensure that $\ket{k}$ is indeed created as desired. The optimization strategy and results will be discussed thoroughly in the next section. For larger systems, one can resort to a variational quantum eigensolver (VQE) to perform energy minimization in each sector using a quantum computer. Such details will be presented in Sec.~\ref{sec:3}. 

Once the optimized $b_k^\dagger$ in each momentum sector is obtained, the wave-packet creation operator $b_\Psi^\dagger$ is just a weighted assembly of them following $\Psi(k)$. Since the simulation is eventually done in position space, it is useful to express Eq.~\eqref{eq:ansatz1} in terms of position-space mesonic operators when implemented as quantum circuits:
\begin{equation}
    b_\Psi^\dagger = \sum_{m,n  \in \Gamma} C_{m,n} \, \widetilde{\mathcal{M}}_{m,n}.
    \label{eq:ansatz4}
\end{equation}
Here, $\widetilde{\mathcal{M}}_{m,n}$ is the Jordan-Wigner transformed $\mathcal{M}_{m,n}$ that is obtained by substituting Eq.~\eqref{eq:jw} in the expression for $\mathcal{M}_{m,n}$ given below Eq.~\eqref{eq:ansatz1}. For example, consider $m<n$ and $n-m < N/2$, then a forward-wrapped meson creation operator $\mathcal{M}_{m,n}$ leads to 
\begin{align}
\widetilde{\mathcal{M}}_{m,n} = \left( \sigma_m^- \sigma_n^+\prod_{l=m+1}^{n-1}\sigma_l^{\textbf{z}}  \right) \otimes \left(\prod_{l=m}^{n-1} U_l\right).
\label{eq:M-tilde}
\end{align}
The coefficient $C_{m,n}$ can be directly extracted from the optimized ansatz function $\eta(p,q)$ and the wave-packet profile $\Psi(k)$:
\begin{align}
C_{m,n}=\sum_k \Psi(k)\sum_{p,q \in \widetilde{\Gamma}} \delta_{k, p+q} &\eta(p, q)
\sqrt{\frac{m_f+\omega_p}{2\pi\omega_p}} \sqrt{\frac{m_f+\omega_q}{2\pi\omega_q}}\nonumber\\
&\times {e^{i(pm+qn)}}\left(\mathcal{P}_{m0}+v_p\mathcal{P}_{m1}\right) \left(-v_q\mathcal{P}_{n0}+\mathcal{P}_{n1}\right).
\label{eq:cmn-def}
\end{align}
Figure~\ref{fig:double_smearing} presents a schematic picture of the wave-packet construction that includes two smearings: the wave-packet profile $\Psi(k)$ with tunable input, and the mesonic ansatz $\eta(p,q)$ to be optimized.

We end this section with a remark. The wave-packet creation operator in Eq.~\eqref{eq:ansatz4} can be generalized to non-Abelian gauge theories in 1+1 D. Let us consider $SU(N)$ LGTs in which fermions transform in the fundamental representation of the gauge group, as in the Standard Model, and further consider only one flavor of fermions as in this work. Then, the only changes in the ansatz are to take into account the multi-component nature of the fermions (hence a generalized Jordan-Wigner transform needs to be applied~\cite{Davoudi:2022xmb,Raychowdhury:2019iki}) and to account for the multi-component nature of link operators that are now non-commuting matrices, hence their order of operation in the ansatz matters. Mesons are still states that carry no gauge-group charges (i.e., colors). There will be more possibilities for building mesons using different components (colors) of fermions matched to the appropriate components (color indices) of the gauge links. Nonetheless, the general form of the ansatz in this section should still apply, although a more elaborate form for the internal wave-function profiles may be needed. We leave developing a detailed ansatz for mesonic excitations in non-Abelian theories in 1+1 D to future work, and here we proceed with testing the fidelity of such an ansatz for the case of the $Z_2$ LGT in 1+1 D. To keep the discussion concise, similar fidelity tests for the $U(1)$ LGT are presented in Appendix~\ref{app:u1-ansatz}. Outlook for generalizing the ansatz to higher-dimensional theories and other gauge groups is further discussed in Sec.~\ref{sec:summary}.

\subsection{Numerical verification of the ansatz in small systems
\label{sec:2-3}}
In this section, we benchmark the wave-packet ansatz against exact results for the $Z_2$ LGT coupled to one flavor of staggered fermions in 1+1 D. Our hardware results for wave-packet preparation in the next section are limited to the case of a 6-site theory due to computational cost. Therefore, we focus on demonstrating the goodness of the optimized ansatz for small systems (6 and 10 sites) in this section. For the $U(1)$ LGT, verification of the ansatz's performance, similarly for a range of model parameters, is presented in Appendix~\ref{app:u1-ansatz}. 
\begin{figure}[t!]
    \centering
    \includegraphics[width=\textwidth]{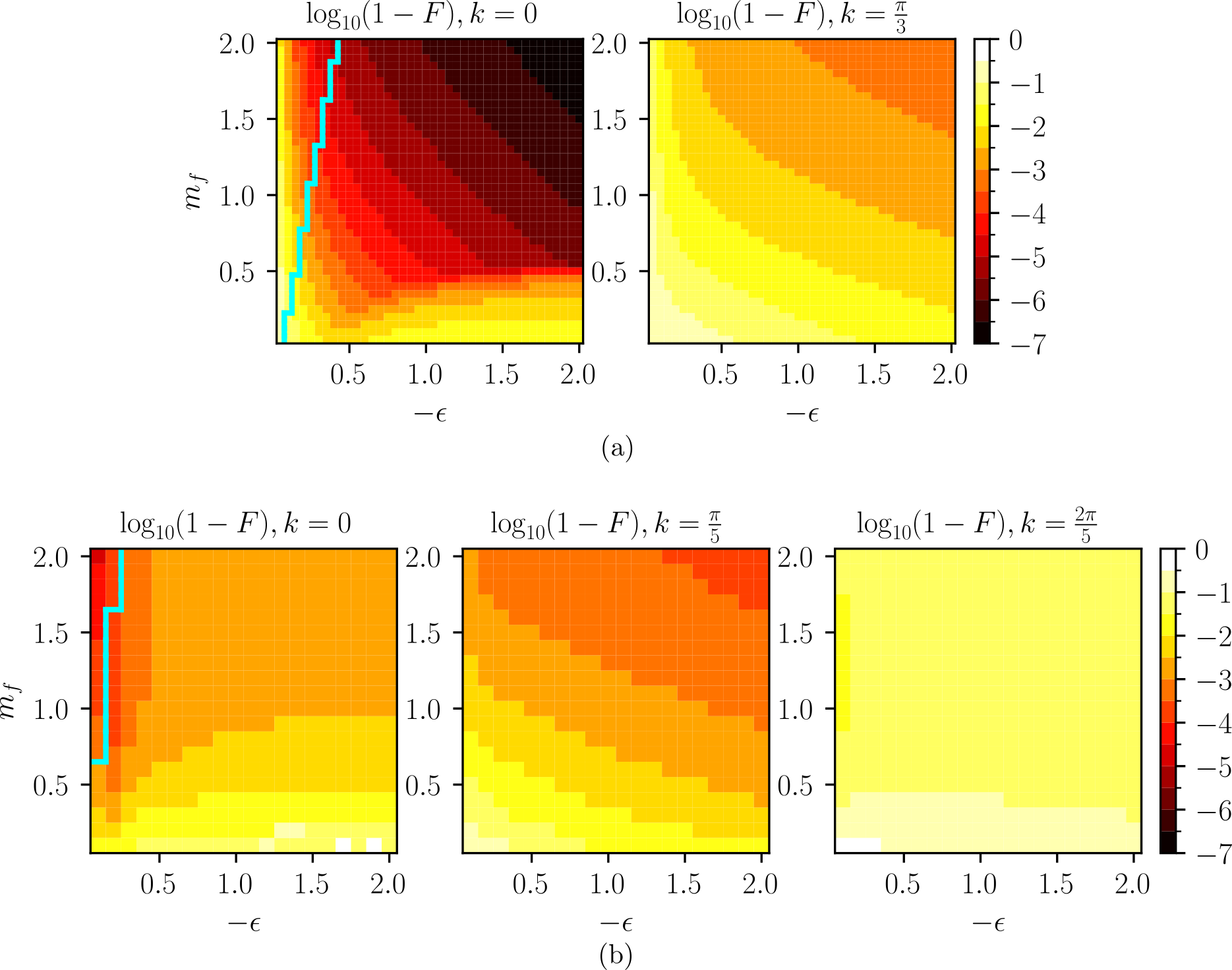}
    \caption{
    The $k$-optimization heat maps for the $Z_2$ LGT. Panels (a) and (b) refer to the 6-site and the 10-site theories, respectively. The region left to the cyan contour is where the non-mesonic excitations are less energetic than the single-meson $\ket{k}=0$ state. For the 6-site theory, the results for $\ket{k=-\frac{\pi}{3}}$ are identical to those for $\ket{k=\frac{\pi}{3}}$ and are not shown. Similarly, the optimization results for $k$ and $-k$ values are identical in the 10-site theory and only positive momentum eigenstates are presented. The ansatz works better in the strong coupling regime, and for momenta closer to the center of the Brillouin zone ($k=0$).}
    \label{fig:scan}
\end{figure}

For a lattice with 6 (10) staggered sites, the Brillouin zone contains 3 (5) discrete momentum sectors, $k\in \widetilde{\Gamma}=\{-\frac{\pi}{3},0,\frac{\pi}{3}\} (\widetilde{\Gamma}=\{-\frac{2\pi}{5},-\frac{\pi}{5},0,\frac{\pi}{5},\frac{2\pi}{5}\})$. So one needs to optimize the ansatz for $b_k^\dagger$ for these $k$ values. First, we examine whether high fidelities can be achieved for $k$-momentum-state optimization with different model parameters. The fidelity of each $k$-momentum state is defined as:
\begin{equation}
    F = {\left|\,{}_{\text{optimized}}\langle k|k\rangle_{\text{exact}}\right|}^2,\label{eq:fidelity}
\end{equation}
where the subscript exact stands for exact-diagonalized momentum eigenstate $\ket{k}$. The fidelity results of scanning $(m_f, \epsilon)$ parameter pairs are shown in Fig.~\ref{fig:scan}. The optimization performs best in the strong-coupling regime for both lattices. It is also able to reach a fidelity $>0.95$ in the intermediate coupling regime (where the energy partition for the hopping term is comparable to the mass term and the electric-field term), except for $k=\pm\frac{2\pi}{5}$ in the 10-site theory. There, the fidelity between the optimized state and the exact-diagonalized state is around 0.90 for most of the parameter space. It is generally observed that the $k$-momentum optimization becomes less accurate with $k$ closer to the boundary of the Brillouin zone, $\widetilde{\Gamma}$. Therefore, if the wave packet is centered further away from the edge of the Brillouin zone, the Gaussian profile of the wave packet will suppress the effect of these lower-fidelity high-momentum states in the full wave-packet state. The parameters $m_f=1$ and $\epsilon=-0.3$ in the 6-site $Z_2$ LGT, used in the hardware demonstration in the next section, correspond to $F>0.98$ for all $k$.

\begin{figure}
    \centering
    \includegraphics[scale=0.95]{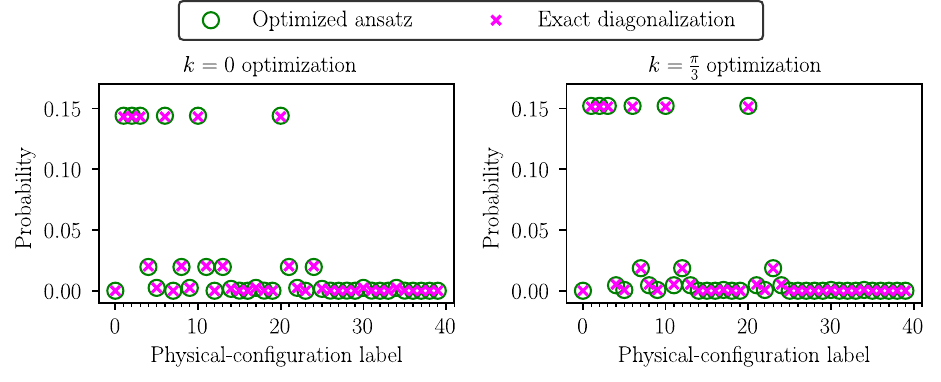}
    \caption{Optimization results for $k=0~(\frac{\pi}{3})$ momentum eigenstates are shown in the left (right) panel for the case of the 6-site $Z_2$ LGT with $m_f=1$ and $\epsilon=-0.3$. The physical basis-state probabilities the of $k=-\frac{\pi}{3}$ state are identical to $k=\frac{\pi}{3}$ and are not shown. The physical basis states of the 6-site $Z_2$ theory are listed in Tables~\ref{tab:physical_z2}.}
    \label{fig:k_optimization}
\end{figure}

The ansatz creation operator is capable of creating mesonic excitations on top of the interacting ground state $\ket{\Omega}$. However, not all states in the interacting spectrum can be generated by such operators. Under PBCs, there are other non-mesonic excitations generated by operators made up of only the gauge-link operators spanning over the entire lattice, e.g., $\prod_{n=0}^{N-1}U_n$ and its conjugate. As a result, the energies of these non-mesonic excitations grow with $|\epsilon|$ and the system size $N$. Furthermore, excitations consisting of only these non-mesonic operators have momentum $k=0$ since these operators are manifestly invariant under translation. In large lattices, states generated from non-mesonic excitations have larger energy compared to the mesonic excitations captured by the ansatz of this work. However, for smaller systems and sufficiently small $|\epsilon|$ values, the $k=0$ sector contains states generated by purely non-mesonic operators that have energies comparable or smaller than the states consisting of purely mesonic excitations.
Such a transition in the Hamiltonian parameter space is denoted by the cyan contour for the $k=0$ heat maps in Fig.~\ref{fig:scan}. In the region left (right) of the cyan contour, the first excited state is created by the purely non-mesonic (mesonic) operators. As expected, this region in the 10-site theory is smaller than that in the 6-site theory. Importantly, our ansatz captures the mesonic states with good fidelity in both regions.


Second, choosing the parameters $m_f=1$ and $\epsilon=-0.3$ as in the next section for the case of $Z_2$, we plot in Fig.~\ref{fig:k_optimization} the probability distribution of each of the physical basis states (called `physical configurations' throughout) in the $Z_2$ theory, comparing those obtained from the optimized ansatz and exact diagonalization. It is seen that each $k$-momentum state created by the ansatz recovers the exact-diagonalized state probabilities. 


\section{Preparing mesonic wave packets in $Z_2$ LGT in 1+1 D
\label{sec:3}
}
\noindent
With high accuracies reached in momentum-eigenstates optimization, in this section, we proceed to construct, and investigate the quality of, the circuits that will generate the full mesonic wave packets.

\subsection{
Algorithm and circuit design
\label{sec:3-1}}
Our wave-packet preparation algorithm, as summarized in the right panel of Fig.~\ref{fig:schematic}, involves four steps: the ground(vacuum)-state preparation, the $k$-momentum-eigenstate optimization, and the wave-packet assembly and circuit implementation. This section details the algorithm for the above components and demonstrates the design of the associated quantum circuits. An important aspect of the circuit is to limit the resources required to fit into Noisy Intermediate-Scale Quantum (NISQ) hardware. With this in mind, we prepare the $Z_2$ interacting ground state, $\ket{\Omega}$, and excited states, $\ket{k}$, based on a VQE, a class of hybrid classical-quantum algorithms suitable especially for the NISQ era of computing~\cite{Peruzzo:2013bzg, McClean:2015vup,Tilly:2021jem}. Applications of VQE-based algorithms to gauge theories have been extensively explored in recent years for ground-state properties and dynamics~\cite{Kokail:2018eiw, Lumia:2021tpu, Farrell:2023fgd, Paulson:2020zjd, Atas:2021ext,Popov:2023xft,Liu:2021otn,Nagano:2023uaq,Xie:2022jgj}. Once $\ket{\Omega}$ and $\ket{k}$ are obtained, a wave-packet preparation circuit can be derived and attached to the ground-state preparation circuit. We list the complete state-preparation algorithm below:
\sloppy
\begin{enumerate}
    \item Define a ground-state preparation circuit $\mathcal{Q}_{\text{GS}}(\theta_i)$ parameterized by a set of single- or multi-qubit rotations $\theta_i$. $\mathcal{Q}_{\text{GS}}$ acts on some initial state $\ket{\psi}_0$ and returns $\mathcal{Q}_{\text{GS}}(\theta_i)\ket{\psi}_0=\ket{\psi(\theta_i)}$.
    Perform VQE using $\mathcal{Q}_{\text{GS}}$ to minimize $\bra{\psi(\theta_i)}H\ket{\psi(\theta_i)}$ and arrive at parameters $\theta_i^{*}$, with $\ket{\psi(\theta_i^{*})}$ obtaining a sufficiently close state to the interacting ground state, i.e., the vacuum $\ket{\Omega}$.
    \item In each $k$-momentum sector, define the circuits $\mathcal{Q}_{\text{k}}(\sigma_k^A,\mu_k^A), \forall k\in\widetilde{\Gamma}$, parameterized by the Gaussian ansatz in Eq.~\eqref{eq:ansatz1}, which is attached to the optimized $\mathcal{Q}_{\text{GS}}(\theta_i^{*})$. Assuming the ground state is prepared accurately, the output would be $\mathcal{Q}_{\text{k}}(\sigma_k^A,\mu_k^A)\ket{\Omega} = \ket{k(\sigma_k^A,\mu_k^A)}$
    Apply VQE to optimize the circuit $\mathcal{Q}_{\text{k}}(\sigma_k^A,\mu_k^A)$ by minimizing $\bra{k(\sigma_k^A,\mu_k^A)}H\ket{k(\sigma_k^A,\mu_k^A)}$ in each $k$-momentum sector.
    Obtain the optimized parameters $(\sigma_k^{A*},\mu_k^{A*})$, with $\ket{k(\sigma_k^{A*},\mu_k^{A*})}$ approximating $\ket{k}$.
    \item In position space, calculate the coefficients $C_{m,n}$ in Eq.~\eqref{eq:ansatz4} by convoluting the wave-packet profile $\Psi(k)$ with the optimized ansatz function $\eta(p,q;\sigma_k^{A*},\mu_k^{A*})$.
    \item Define the wave-packet preparation circuit $\mathcal{Q}_{\text{WP}}(C_{m,n})$, which is attached to the optimized $\mathcal{Q}_{\text{GS}}(\theta_i^{*})$. Output $\mathcal{Q}_{\text{WP}}(C_{m,n})\mathcal{Q}_{\text{GS}}(\theta_i^{*})\ket{\psi_0}=\mathcal{Q}_{\text{WP}}(C_{m,n})\ket{\Omega}=\ket{\Psi}$.
\end{enumerate}

We emphasize that this is a hybrid classical-quantum algorithm, with both the ground-state and $k$-momentum-eigenstate preparation being obtained by VQE. In this work, VQE for both the ground-state and the $k$-momentum eigenstates (steps 1 and 2) are only verified through classical evaluation. This is solely due to efficiency considerations and resource limitations. 
In principle, one can implement both the ground-state and the $k$-momentum VQEs provided that the number of measurement shots is sufficient to resolve the energy.
\begin{figure}
    \centering
    \includegraphics[scale=0.95]{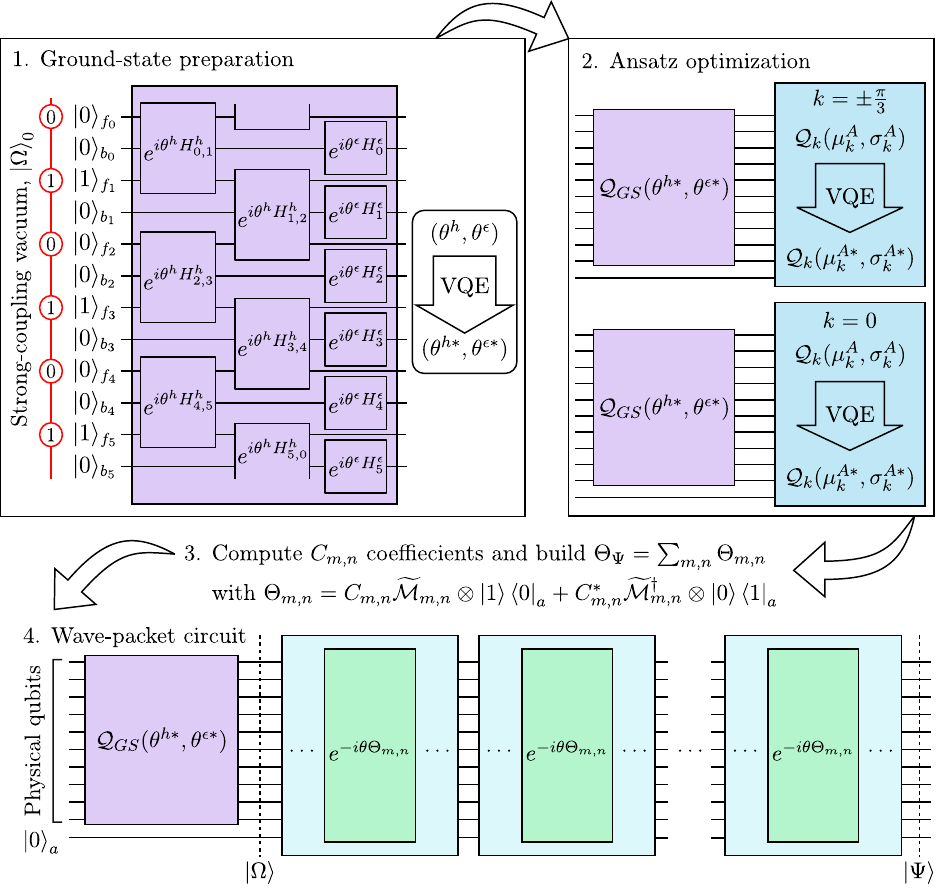}
    \caption{ 
    The schematic diagram of the full algorithm. The numbers correspond to the steps described in Sec.~\ref{sec:3-1}. The purple block in step 1 depicts the circuit for $\mathcal{Q}_{\rm GS}$ parameterized by $\left(\theta^{h},\theta^{\epsilon} \right)$ with $N_{\rm GS}=1$, see Eq.~\eqref{eq:algo_gs_II}.
    The evolution operators corresponding to the hopping (electric-field) terms in the Hamiltonian are denoted by multi-(single-)qubit gates.
    Qubits representing fermions and bosons are labeled by $f$ and $b$, respectively, and the subscripts are the site indices.
    The qubit mapping of the initial state $\ket{\Omega}_0$ is shown next to its basis-state representation.
    VQE is applied to find the optimum parameters $\left(\theta^{h*},\theta^{\epsilon*} \right)$ that minimize the energy with respect to the $Z_2$ LGT Hamiltonian. The blue block in the top (bottom) circuit in step 2 denotes the circuit for $\mathcal{Q}_{k}$ parameterized by the ansatz parameters $(\sigma_k^A, \mu_k^A)$, for $k=\pm \frac{\pi}{3}$ ($k=0$). VQE is employed to obtain the parameters $(\sigma_k^{A*}, \mu_k^{A*})$ that optimize the ansatz for each $k$.
    The full circuit that prepares a wave packet $\ket{\Psi}$ from $\ket{\Omega}_0$ is shown in step 4, with the ancilla qubit labeled by $a$. The repeated blue blocks acting on $\ket{\Omega}$ denote the $\mathcal{Q}_{\rm WP}(C_{m,n})$ circuit with each blue block implementing one Trotter step in the Trotterized Eq.~\eqref{eq:anc1}. The green block indicates the SVD circuit for $e^{-i\theta \Theta_{m,n}}$ and is described in detail in Fig.~\ref{fig:circuit}.
    }
    \label{fig:full-circuit}
\end{figure}
We discuss the various steps of the algorithm in detail in the following.

\subsubsection{Ground-state preparation}
\label{sec:3-1-1}
Here, we demonstrate a VQE-based ground-state preparation method that meets the requirement of a shallow circuit with low gate counts. We follow the general strategy in Ref.~\cite{Lumia:2021tpu} where the desired entanglement of the ground state is built in by evolving some initial state with operators that respect the symmetry of the theory. The natural candidates for the evolution operators are, therefore, the hopping term, the mass term, and the electric-field term in the Hamiltonian (labeled $H^h = \sum_{n \in \Gamma} H_{n,n+1}$, $H^m=\sum_{n \in \Gamma}H^m_n$, and $H^\epsilon=\sum_{n \in \Gamma} H^\epsilon_n$ below, respectively). The initial state $\ket{\psi}_0$ is chosen to be the strong-coupling vacuum $\ket{\Omega}_{0}$. Explicitly,
\begin{equation}
\ket{\Omega} = \mathcal{Q}_{\text{GS}} \ket{\Omega}_{0},
\label{eq:algo_gs_I}
\end{equation}
with 
\begin{equation}
\mathcal{Q}_{\text{GS}} = \prod_{j=1}^{N_{\rm GS}} \left(\prod_{n \in \Gamma}{e^{i \theta_j^{h} H_{n,n+1}^{h} }}\right) \left(\prod_{n \in \Gamma}e^{i \theta_j^{m} H^{m}_n }\right) \left(\prod_{n \in \Gamma }e^{i \theta_j^{\epsilon} H^{\epsilon}_n }\right).
\label{eq:algo_gs_II}
\end{equation}
In practice, it is found that evolving the hopping term and the electric-field term for only one step, i.e., $N_{\rm GS}=1$, is enough to create $\ket{\Omega}$ with high precision for the lattice size used in this work. One can thus identify $\left(\theta^{h},\theta^{\epsilon} \right)$ as the circuit parameters and perform optimization as instructed. The detailed circuit 
is depicted in Fig.~\ref{fig:full-circuit}. The quality of $\ket{\Omega}$ 
arising from this VQE ansatz and optimization will be discussed in Sec.~\ref{sec:4-1}. The cost of ground-state preparation only depends on the system size as $H_{n,n+1}^{h}$ is a nearest-neighbor term, effectively a 1-length meson. 
If the VQE demands $N_{\rm GS}>1$ layers, the CNOT-gate cost gets multiplied by $N_{\rm GS}$.

Next, VQE can be applied to obtain the optimized values of $(\sigma_k^A, \mu_k^A)$, hence determining a good approximation to each $\ket{k}$. This involves applying the full wave-packet circuit $\mathcal{Q}_{\text{WP}}$ with the outer wave-packet profile set to a delta function in momentum space. This circuit is denoted as $\mathcal{Q}_k$ in Fig.~\ref{fig:full-circuit}. As a result, the $k$-momentum-eigenstate preparation has an identical circuit construction to the full wave-packet circuit, amounting to realizing the $b_\Psi^\dagger$ operator in Eq.~\eqref{eq:ansatz4}. Hence in the following, we will focus on demonstrating the elements to build such a circuit.

There are two issues with regards to circuitizing $b_\Psi^\dagger$. First, note that the $b_\Psi^\dagger$ operator is not unitary, and thus requires a strategy to implement it via a unitary quantum circuit. Furthermore, each mesonic creation operator [summand in Eq.~\eqref{eq:ansatz4}] will be a multi-spin operator, and the summation circuit can be tedious to construct and implement. A suitable circuit implementation is required to keep the number of entangling gates within NISQ-hardware limitations. We will address these aspects of the quantum-circuit construction in the following.

\subsubsection{Unitary implementation of creation operators}
\label{sec:3-2-2}
To implement $\mathcal{Q}_{\text{WP}}$, one can embed the non-unitary creation operators within a unitary operator that acts on an extended Hilbert space. In this work, the ancilla encoding introduced in Ref.~\cite{Jordan:2011ci} is used to realize such operators. Consider a non-unitary creation operator $b_\Psi^{\dagger}$. One can introduce an ancilla qubit, and define the following Hermitian operator:
\begin{align}
\Theta_{\Psi} =  b_\Psi^{\dagger} \otimes \ket{1}\bra{0}_{a} +  b_\Psi \otimes \ket{0}\bra{1}_{a},
\end{align}
with subscript $a$ representing the ancilla. It is then easy to show that 
\begin{equation}
e^{-i \frac{\pi}{2} \Theta_{\Psi}} \ket{\Omega} \otimes \ket{0}_a = -i \ket{\Psi} \otimes \ket{1}_a,
\label{eq:anc1}
\end{equation}
provided that $b_\Psi \ket{\Omega}=0$ and the commutation relation $ [b_\Psi, b_\Psi^{\dagger}] = \mathds{1}$ holds. These conditions are met for the $b_\Psi$ operator considered in Eqs.~\eqref{eq:bk3}. This method, therefore, provides a straightforward realization of such non-unitary creation operators on quantum circuits with a minimal number of ancillary qubits. 

As mentioned in Sec.~\ref{sec:2-2}, the creation operator $b_\Psi^\dagger$ is a linear combination of many position-space mesonic creation operators with coefficients $C_{m,n}$, see Eq.~\eqref{eq:ansatz4} and subsequent equations. In general, these mesonic operators do not commute, and to circuitize the full creation operator, a Trotterization scheme is required. Nonetheless, the method above still works, up to a trotter error~\cite{Childs:2019hts}, as long as the resulting $\ket{\Psi}$ is properly normalized.

\subsubsection{Singular-value-decomposition circuit}
\label{sec:3-3}
Given the form of $b_\Psi^\dagger$ in Eq.~\eqref{eq:ansatz4}, one can define
\begin{align}
\Theta_\Psi=\sum_{m,n}\Theta_{m,n},
\end{align}
with
\begin{align}
\Theta_{m,n} =  C_{m,n}\widetilde{\mathcal{M}}_{m,n} \otimes \ket{1}\bra{0}_{a} +  C_{m,n}^*\widetilde{\mathcal{M}}_{m,n}^\dagger \otimes \ket{0}\bra{1}_{a}.
\label{eq:Theta-m-n}
\end{align}
The goal is to implement $e^{-i\theta \Theta_{m,n}}$ for $\theta \in \mathbb{R}$. Each $\Theta_{m,n}$ is a multi-qubit operator in general, see e.g., Eq.~\eqref{eq:M-tilde}. A noteworthy structure of the multi-qubit operators in $\Theta_{m,n}$ for $m \neq n$~\footnote{For $m=n$, the implementation of $e^{-i\theta\Theta_{m,n}}$ reduces to a simple two-qubit rotation.} is that they appear as $A^{\dagger} \otimes\ket{1}\bra{0}_a+ A\otimes\ket{0}\bra{1}_a$.
On top of that, these $A$ and $A^\dagger$ operators satisfy $A^2 = {A^\dagger}^2 = 0$ as a result of their fermionic content. According to Ref.~\cite{Davoudi:2022xmb}, a circuit realizing the exponentiation of operator forms can be efficiently constructed by finding a diagonal basis via a singular-value decomposition (SVD) of $A$, provided that this SVD is easy to identify. The procedure goes as follows:
\begin{enumerate}[i)]
    \item For an operator $\Theta = A^{\dagger} \otimes \ket{1}\bra{0}_a + A \otimes \ket{0}\bra{1}_a$, identify the SVD, $A = VSW^\dagger$, where $S$ is a real diagonal matrix. The subscript $a$ denotes the ancilla-qubit space. 
    \item Diagonalize the operator $\Theta = \mathcal{U}^\dagger \mathcal{D} \mathcal{U}$ with $\mathcal{U} = \mathsf{H}_a( V^\dagger \otimes {\ket{0}\bra{0}}_a + W^\dagger \otimes {\ket{1}\bra{1}}_a) $ and diagonal $\mathcal{D} = S \otimes \sigma^\textbf{z}_a$. $\mathsf{H}=\frac{1}{\sqrt{2}}\big(\begin{smallmatrix}
    1 & 1\\
    1 & -1
    \end{smallmatrix}\big)$ stands for the Hadamard gate. ${\ket{0}\bra{0}}_a$ and ${\ket{1}\bra{1}}_a$ are the projection operators that can be realized by controlled gates.  
    \item Plug in the form of $V$, $S$, and $W$ from the SVD into $\mathcal{U}$ and $\mathcal{D}$. Implementing $e^{-i\theta\Theta}$ is then equivalent to implementing $\mathcal{U}^{\dagger} e^{-i\theta\mathcal{D}} \mathcal{U}$ instead.
\end{enumerate}
\begin{figure}[t!]
    \centering
    \includegraphics[width=\textwidth]{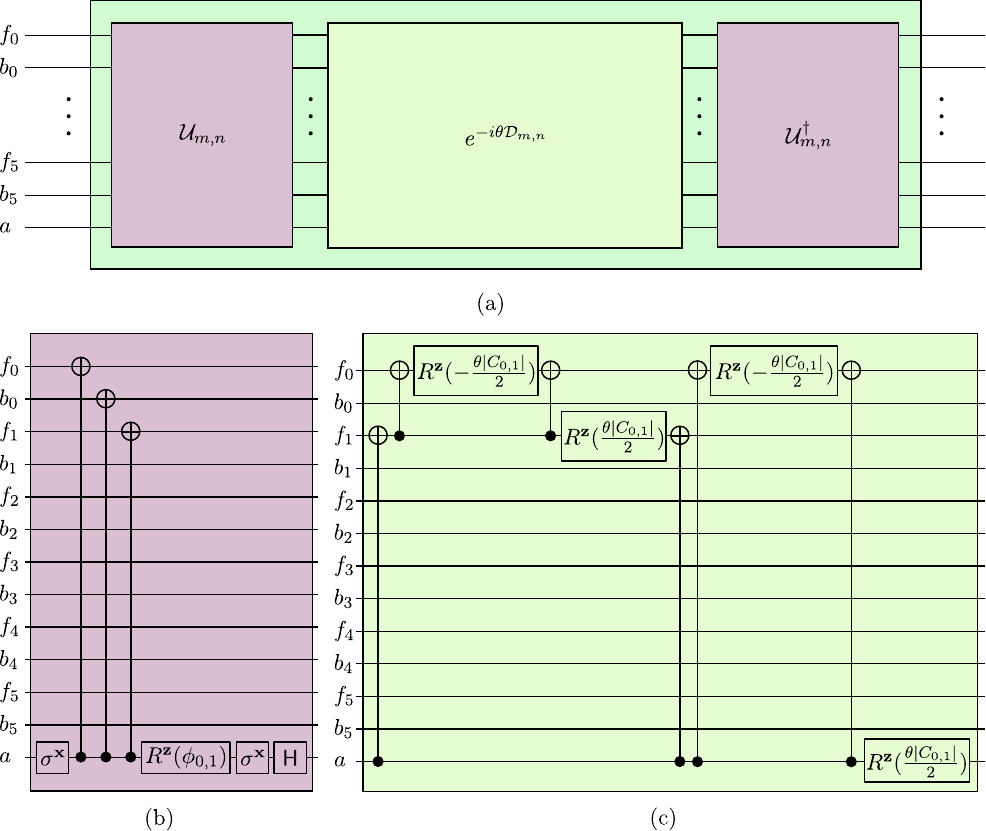}
    \caption{The dark green block in (a) shows the SVD circuit implementing $e^{-i\theta\Theta_{m,n}}$ in $\mathcal{Q}_{WP}$ on a 6-site system.
    The circuits for $\mathcal{U}_{m,n}$ and $\mathcal{U}^\dagger_{m,n}$ that diagonalize $e^{-i\theta\Theta_{m,n}}$ are denoted by pink blocks, and the circuit for diagonalized $e^{-i\theta\mathcal{D}_{m,n}}$ is represented by the light-green block.
    As an example, the circuit blocks for $\mathcal{U}_{m,n}$ and $e^{-i\theta\mathcal{D}_{m,n}}$ for the case $m=0$, $n=1$ are expanded according to their forms in Eqs.~\eqref{eq:svd_U} and \eqref{eq:svd_D}, respectively, and shown here in terms of single-qubit and CNOT gates in (b) and (c).
    } 
    \label{fig:circuit}
\end{figure}

Consider the creation operator summand in Eq.~\eqref{eq:ansatz4} and its ancilla encoding, $\Theta_{m,n}$. As an example, consider $m<n$ and $|m-n|<N/2$ so that the operator has the form $C_{m,n} \sigma_m^- \sigma_n^+ \left(\prod_{l=m+1}^{n-1}{\sigma_l^{\textbf{z}}}\right)$ $\left(\prod_{l=m}^{n-1}{\tilde{\sigma}_l^{\textbf{x}}}\right)$. Further, let $ C_{m,n}=e^{i\phi_{m,n}}\left|C_{m,n}\right|$. Then one can identify the $V$ and $W$ matrices as: 
\begin{subequations}
\begin{align}
    &V_{m,n} =  e^{\frac{-i\phi_{m,n}}{2}}\sigma_m^{\textbf{x}}\sigma_n^{\textbf{x}}\left(\prod_{l=m}^{n-1}{\tilde{\sigma}_l^{\textbf{x}}}\right),
    \label{eq:svd_V}\\
    &W_{m,n} =  e^{\frac{i\phi_{m,n}}{2}}\mathds{1}.
    \label{eq:svd_W}
\end{align}
\end{subequations}
With the aid of the unitary:
\begin{align}
    &\mathcal{U}_{m,n} = \mathsf{H}_a \sigma_a^\textbf{x} R^{\bf{z}}_a(\phi_{m,n}) \text{C}^{\text{x}}_{a,m}\text{C}^{\text{x}}_{a,n} \left(\prod_{l=m}^{n-1}{\text{C}^{\tilde{\text{x}}}_{a,l}}\right)\sigma_a^\textbf{x},
    \label{eq:svd_U}
\end{align}
the intended operator is diagonalized as $\mathcal{U}^\dagger \mathcal{D} \mathcal{U}$ with
\begin{align}
    &\mathcal{D}_{m,n} = \left|C_{m,n}\right| 
    \left(\frac{\mathds{1}_m-\sigma_m^\textbf{z}}{2}\right)\left(\frac{\mathds{1}_n+\sigma_n^\textbf{z}}{2}\right)
    \left(\prod_{l=m+1}^{n-1}\sigma_l^{\textbf{z}}\right)\otimes\sigma_a^{\textbf{z}}.
    \label{eq:svd_D}
\end{align}
Here, $\text{C}^{\text{x}}_{i,j}$ is the CNOT gate with control qubit $i$ and target qubit $j$, $\text{C}^{\tilde{\text{x}}}_{i,j}$ denotes the same operation when the target qubit corresponds to the gauge-field degrees of freedom, and ${R^\textbf{z}}(\phi)=e^{-i\phi \sigma^\textbf{z}/2}$ is the single-qubit $z$-rotation gate. Compared to the naive Pauli decomposition of the $e^{-i\theta \Theta}$ operator, the SVD algorithm turns out to reduce the number of entangling gates.

Examples of circuit elements are explicitly shown in Fig.~\ref{fig:circuit}. The number of entangling gates in the $\mathcal{Q}_\text{WP}$ circuit depends on the meson length $|m-n|$, which determines the number of CNOT gates in $\mathcal{U}$ and $\mathcal{U}^\dagger$, as well as the size of the Jordan-Wigner Pauli string in $\mathcal{D}$. As discussed above, the wave-packet creation operator requires Trotterization due to non-commuting summands of mesonic operators. The Trotterization order, $N_{\text{Order}}$, and the number of Trotter steps, $N_{\text{Trotter}}$, are, therefore, introduced as circuit parameters. We consider a second-order product formula ($N_{\text{Order}}=2$) throughout this work, and the number of Trotter steps, $N_{\text{Trotter}}$, contributes multiplicatively toward the total resource counting. Furthermore, when the circuit is implemented on the hardware, we consider a truncation, $\theta_{c}$, on the coefficients $|C_{m,n}|$ when implementing $e^{-i\frac{\pi}{2} \Theta_{m,n}}$. 
This turns out to be effectively omitting the creation operators that produce larger mesons. To what extent these large mesons are truncated depends on the parameters of the theory ($m_f$, $\epsilon$) and of the wave packet ($\sigma$, $\mu$ and $k_0$), which will be further discussed in Appendix~\ref{app:truncation}. A general observation is that the distribution of $|C_{m,n}|$ tends to be more uniform across sites $(m,n)$ in the low-mass regime. This implies more resources are required to prepare the state due to a higher surviving chance of the mesonic operators at a given $\theta_{c}$. As will be seen in Sec.~\ref{sec:4-2} and Appendix~\ref{app:truncation}, $\theta_{c}$ is chosen such that mostly 1-length mesons are created. 

The resource counting (number of qubits and entangling gates) of the circuits for ground-state, the full wave packet, and the 1-meson truncated wave-packet preparation are listed in Table~\ref{tabel:gate-count}. 
To further simplify the circuit, we have assumed that $N=4\mathbb{N}-2$ for positive integer $\mathbb{N}$ such that the Jordan-Wigner $\sigma^\textbf{z}$ strings in $\Theta_{m,n}$ are always implemented along the shorter path between $m$ and $n$. This choice is consistent with that from the original Jordan-Wigner transformation Eq.~\eqref{eq:jw}, as long as $(-1)^{Q+1}=1$ (see discussions regarding PBCs and Jordan-Wigner transform in Sec.~\ref{sec:2-1}). Since we have restricted the analysis to the $Q=N/2$ sector, for $N=4\mathbb{N}-2$, such an efficient implementation is justified.
Then for an $N$-site lattice, there are $N^2$ different $(m,n)$ pairs in position space. Recall that the ansatz only creates shorter mesons, and keep the other wrapping only when the forward-wrapped and backward-wrapped mesons have the same length. Thus, one can collect $N$ 0-length mesons and $2N$ $l$-length mesons where $0<l\leq\frac{N}{2}$. The creation operators of 0-length mesons are already diagonal, therefore require no $\mathcal{U}$ and $\mathcal{U}^\dagger$ circuits, and 2 CNOT gate for $e^{-i\theta\mathcal{D}}$. For each $l$-length meson creation operator, the number of entangling gates depends on $l$. In Fig.~\ref{fig:circuit}, one can see that $l+2$ and $2l+4$ CNOT gates are needed for $\mathcal{U}$ and $e^{-i\theta\mathcal{D}}$, respectively. Summing $l$ from 1 to $\frac{N}{2}$ and multiplying by the $2N$ factor, one can recover the $N^3$ dependence in Table~\ref{tabel:gate-count}. 
If the circuit is truncated such that only the $l\leq1$ summands are considered, the gate count for individual operators will be constant, and the scaling of the total CNOT gates will be linear in $N$ as shown in Table~\ref{tabel:gate-count}. In summary, the number of gates required to prepare the wave packet scales at worst as $N^3\times N_{\rm Trotter}$, and can be reduced to $N\times N_{\rm Trotter}$ if only mesons with length less or equal to 1 are implemented.

Overall, the performance of the algorithm is also a function of the performance of the ansatz, which as observed in Fig.~\ref{fig:scan}, depends on the systems parameters $N$, $m_f$, and $\epsilon$. Unfortunately, without the analytical knowledge of the states (as is the case in nonperturbative theories), an analytic bound cannot be found on the performance of the ansatz, hence the algorithm. Our study, nonetheless, establishes empirical evidence of an acceptable algorithm performance in small systems and for a range of parameters.

\begin{table}[]
\centering
\begin{tabular}{|c|c|c|c|c|}
\cline{2-5} 
\multicolumn{1}{c|}{} & Qubit & \multicolumn{3}{c|}{CNOT-gate count}\tabularnewline
\cline{3-5} 
\multicolumn{1}{c|}{} & count & $\mathcal{U}$ & $\mathcal{D}$ & Total\tabularnewline
\hline 
\hline 
Ground-state cicuit & $2N$ & - & - & $6N$\tabularnewline
\hline 
Wave-packet circuit & \multirow{2}{*}{$2N+1$} & \multirow{2}{*}{$\frac{1}{4}N^{3}+\frac{5}{2}N^{2}$} & \multirow{2}{*}{$\frac{1}{2}N^{3}+5N^{2}+2N$} & {$\left(N^{3}+10N^{2}+2N\right)$}\tabularnewline
(full) &  &  &  & $\times 2N_{\textrm{Trotter}}$\tabularnewline
\hline 
Wave-packet circuit & \multirow{2}{*}{$2N+1$} & \multirow{2}{*}{$6N$} & \multirow{2}{*}{$14N$} & \multirow{2}{*}{$\ 26N\times 2N_{\textrm{Trotter}}$}\tabularnewline
(1-meson truncated) &  &  &  & \tabularnewline
\hline 
\end{tabular}
\caption{The number of qubits and entangling (CNOT) gates for ground-state and wave-packet preparation circuits for the example of the $Z_2$ LGT coupled to one flavor of staggered fermions in 1+1 D. Here, $N$ denotes the number of staggered sites. Wave-packet circuits require an ancilla qubit. The number of CNOT gates associated with $\mathcal{U}$ and $e^{-i\theta\mathcal{D}}$, as well as the total CNOT-gate count are provided. The full circuit requires resource that scales with $\mathcal{O(}N^3)$, multiplied by $2N_{\rm Trotter}$ for a second-order Trotterization. The cost for the wave-packet circuit with non-zero $\theta_c$ that truncates mesons larger than length one is also presented. The number of entangling gates is greatly reduced to $\mathcal{O}(N)$ for truncated circuits at the cost of a systematic error associated with dropping `unimportant' mesons with larger length.}
\label{tabel:gate-count}
\end{table}

\subsection{
Circuit implementation and results}
\label{sec:3-2}
In order to demonstrate how the algorithm of this work performs in practice, we present in this section hardware-implementation results for the wave-packet creation on a 6-site system in the $Z_2$ LGT in 1+1 D. These results are compared against the exact states obtained by classical simulations of the wave-packet circuit, referred to as `statevector' evolution in the following. The $Z_2$ LGT parameters are chosen to be $m_f=1$ an $\epsilon=-0.3$, for which $b_k^\dagger$ creates momentum eigenstates with high fidelity (see Sec.~\ref{sec:2-3}). 

For statevector evolution, the resulting wave function is guaranteed to stay in the physical Hilbert space as all elements in the circuit are gauge invariant. For the wave packet created on Quantinuum's hardware, errors can take the simulation outside the physical Hilbert space. We will employ a simple error-mitigation scheme to deal with the gauge-violating error in the hardware. As mentioned before, no VQE circuits were implemented on the hardware given its greater computational cost. Instead, VQE with statevector evolution was used to find the optimized values of the parameters that prepare the interacting ground state. This circuit was then implemented to initialize the full wave-packet simulation on the hardware.

\subsubsection{Ground-state preparation
}
\label{sec:4-1}
\begin{figure}[t!]
    \centering
    \includegraphics[]{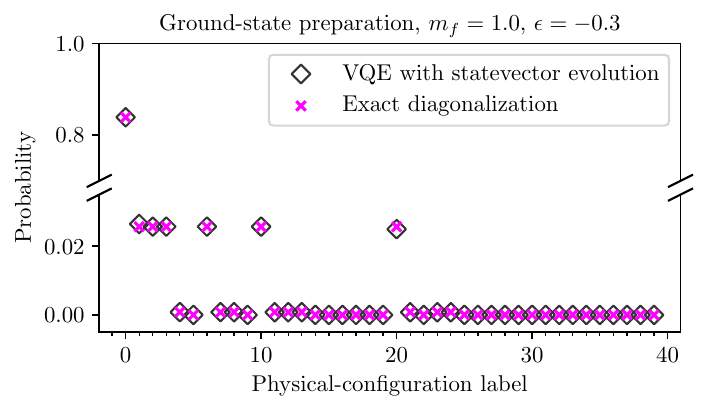}
    \caption{Probabilities of the physical basis states in the ground-state wave function. For 6 sites, there are 40 physical configurations in total. (Classically-simulated) VQE is able to approach the exact-diagonalization results with high accuracy. The $0^{\text{th}}$ state with the highest probability is the strong-coupling vacuum. The basis states are listed in Table~\ref{tab:physical_z2} in Appendix~\ref{app:state-labels}.}
    \label{fig:GS_prob}
\end{figure}
Let us first demonstrate the accuracy of the interacting ground state prepared by $\mathcal{Q}_{\text{GS}}$. Figure~\ref{fig:GS_prob} shows the physical basis-state probabilities in the ground state obtained by exact diagonalization compared with the statevector-evolution results via the VQE circuit proposed. As is observed, the VQE algorithm is able to capture the probabilities accurately. The quality of the result can also be quantified by the fidelity between the output of the optimized circuit, $\ket{\Omega}_{\text{circuit}}$, and the exact-diagonalized ground state, $\ket{\Omega}_{\text{exact}}$,
\begin{equation}
    F_{\rm GS} = {\left|\,{}_{\text{exact}}\langle\Omega|\Omega\rangle_{\text{circuit}}\right|}^2,\label{eq:fidelity_gs}
\end{equation}
as well as by the difference between the energy expectation values using the output state compared with the exact state, $\Delta E_{\rm GS}$. For $m_f=1$ and $\epsilon=-0.3$, we get $1-F_{\rm GS}=7.83 \times 10^{-5}$ and $|\Delta E_{\rm GS}|=3.09 \times 10^{-4}$ (with the exact ground-state energy being $-5.3248$). The associated optimum parameters are $\theta^h \approx -0.34$ and $\theta^\epsilon \approx 15.70$.

\subsubsection{Full wave-packet simulation}
\label{sec:4-2}
Now we consider the full wave-packet creation, which involves attaching the ground-state preparation and the $b_\Psi^\dagger$ circuits, once the parameters of both are found by (classical or quantum) optimization. Ideally, the circuit with $N_{\text{Trotter}}\rightarrow\infty$ and $\theta_c\rightarrow0$ recovers the exact unitary evolution via $\mathcal{Q}_{\text{WP}}$, which creates the desired wave packets. We label the wave-packet states obtained by circuits with $N_{\text{Trotter}}=10$ and $\theta_c=0$ as $\ket{\Psi(\sigma,\mu,k_0)}_\text{ideal}$. For the truncated circuits that only create `important' mesons, $N_{\text{Trotter}}=1$ and $\theta_c=0.1$ are used and the wave-packet state is labeled $\ket{\Psi(\sigma,\mu,k_0)}_\text{trunc}$. This truncated circuit is implemented on the Quantinuum \texttt{H1-1} system. For a more detailed discussion on the effect of $N_{\text{Trotter}}$ and $\theta_c$ on the circuits, see Appendix~\ref{app:truncation}. In summary, we observe that $\ket{\Psi}_\text{trunc}$ approaches  $\ket{\Psi}_\text{ideal}$ quickly as $N_{\text{Trotter}}$ increases. For  $\theta_c=0.1$, $N_{\text{Trotter}}=1$ is sufficient for achieving a fidelity greater than $0.97$. Finally, for small systems, one can directly compute the action of $b_\Psi^\dagger = \sum_k{\Psi(k) b_k^\dagger}$ on $\ket{\Omega}$ without resorting to any circuit decomposition. This is labeled as $\ket{\Psi(\sigma,\mu,k_0)}_{\text{exact}}$ and serves as the benchmark quantity for the circuit-generated states $\ket{\Psi}_{\text{ideal}}$ and $\ket{\Psi}_{\text{trunc}}$. As in the ground-state preparation, we calculate the basis-state probabilities for various $\ket{\Psi}$ as defined above. Furthermore, to verify the spatial properties of the wave packets, the staggered density 
\begin{align}
    &\chi_n = 
    \begin{cases}
        \bra{\Psi}\xi_n^\dagger\xi_n\ket{\Psi}, n\in \text{even} \\
        1-\bra{\Psi}\xi_n^\dagger\xi_n\ket{\Psi}, n\in \text{odd}
    \end{cases} \label{eq:chi1}
\end{align}
is also computed for those states.

\begin{figure}
    \centering
    \includegraphics[width=1\textwidth]{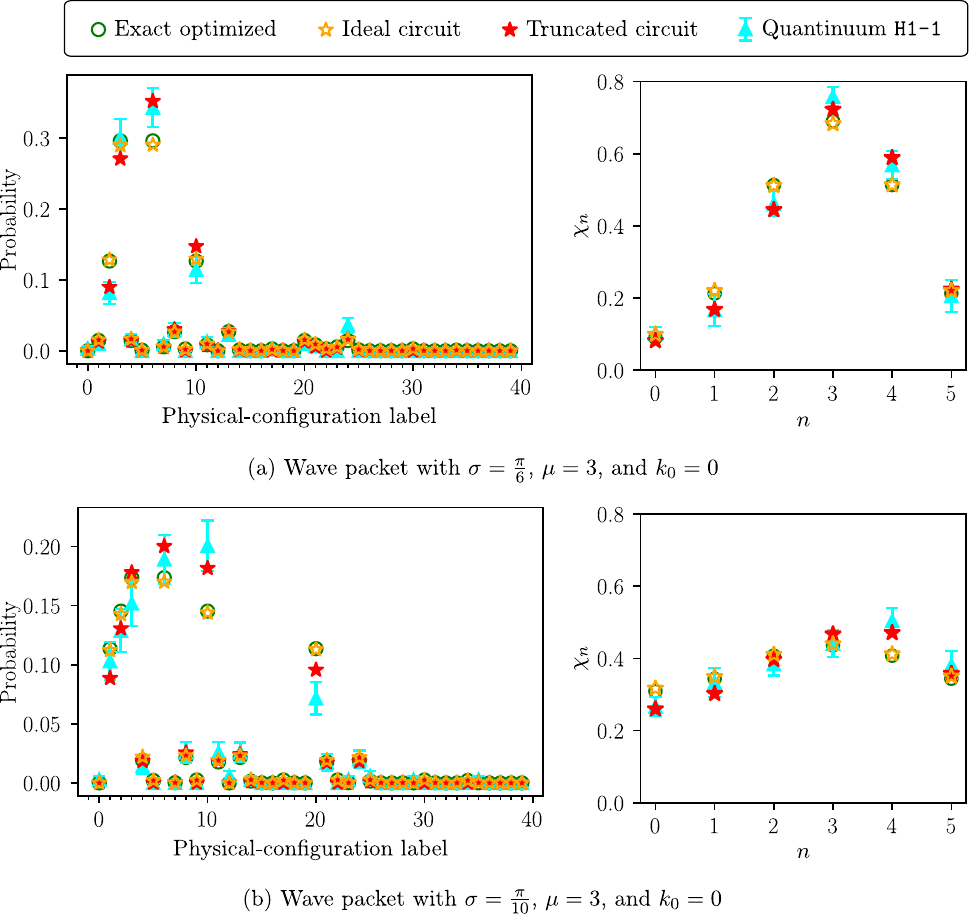}
    \caption{
    Physical basis-state probabilities of two wave packets in the case of $Z_2$ LGT with $m_f=1$ and $\epsilon=-0.3$, generated on the Quantinuum \texttt{H1-1} quantum computer compared with those obtained from $\ket{\Psi}_\text{exact}$, $\ket{\Psi}_\text{ideal}$, and $\ket{\Psi}_\text{trunc}$, which are wave-packet states from exact optimization, and from classical simulation of exact and truncated circuits, respectively, as defined in the text. The associated local particle densities, $\chi_n$ calculated based on these states, along with that obtained from hardware, are presented on the right panels. The physical basis states are listed in Table~\ref{tab:physical_z2} in Appendix~\ref{app:state-labels}. The hardware results shown are after a symmetry-based error mitigation as discussed in the text. These agree reasonably well with the truncated-circuit output. The density plots clearly show the change in the shape of wave packets with varying $\sigma$, i.e., larger $\sigma$ results in a narrower wave packet in position space.
    \label{fig:wp_all}}
\end{figure}

Figure~\ref{fig:wp_all} includes plots for the probabilities in the physical Hilbert space of wave packets with $(\sigma,\mu,k_0)=(\frac{\pi}{6}, 3, 0)$ and $(\sigma,\mu,k_0)=(\frac{\pi}{10}, 3, 0)$, calculated based on $\ket{\Psi}_{\text{exact}}$, $\ket{\Psi}_{\text{ideal}}$, $\ket{\Psi}_{\text{trunc}}$, as well as the hardware results using 500 measurement shots. The number of shots is determined by the relative size of the statistical shot noise compared to other systematic error, as explained in Appendix~\ref{app:shots}. Figure~\ref{fig:wp_all} also includes the staggered density calculated with the two sets of $\left(\sigma,\mu,k_0 \right)$. Let us discuss these results more closely.

First note that there is a small but visible difference between the exact results and the ideal-circuit outputs. This discrepancy is attributed to a systematic error that originates from the unitary encoding in Sec.~\ref{sec:3-2-2}. To ensure that the state $\ket{\Psi}$ gets properly encoded, the conditions $b_\Psi \ket{\Omega}=0$ and $\left[b_\Psi, b_\Psi^\dagger\right]=\mathds{1}$ need to hold. A small violation of these conditions, arising from small difference between optimized $b_k^\dagger$ and the exact ones, results in imperfect encoding of $\ket{\Psi}$ using unitary operations. For an ideal creation operator ${b_k^\dagger}$ that satisfies the conditions stated, the circuit will prepare the wave-packet state assembled with $\Psi(k)\ket{k}$.

Perhaps the largest difference among the probabilities in Fig.~\ref{fig:wp_all} is associated with those obtained from $\ket{\Psi}_\text{ideal}$ and $\ket{\Psi}_\text{trunc}$. This is expected, as keeping mostly length-1 mesons introduces a systematic error. This uncertainty is purely driven by hardware limitations and can be removed once the decoherence time of the quantum hardware is improved. For comparison, implementing the full $\mathcal{Q}_{\rm WP}$ circuit amounts to applying $3192 \times 2N_{\rm Trotter}$ entangling (CNOT) gates, while the circuit truncated with $\theta_c=0.1$ only requires $136\times 2N_{\rm Trotter}$ CNOT gates for $\sigma_k=\frac{\pi}{6}$ [panel (a) of Fig.~\ref{fig:wp_all}] and $72\times 2N_{\rm Trotter}$ for $\sigma=\frac{\pi}{10}$ [panel (b) of Fig.~\ref{fig:wp_all}]. Truncating the circuit, along with the above-mentioned error from the unitary encoding, result in a non-zero probability of measuring 0 on the ancilla. The probabilities shown in Fig.~\ref{fig:wp_all} are obtained with the ancilla measured to be 1, followed by a normalization such that they are summed up to 1. 

Finally, the hardware results is contaminated by various sources of error prevalent to trapped-ion quantum devices~\cite{QuantinuumRef}. Our resources in terms of both the number of shots and circuit depth forbid us from implementing proper error-mitigation protocols, such as the Pauli-twirling-based methods used in Ref.~\cite{Farrell:2023fgd}. Still, it is possible to improve the outcome purely by post-processing the machine readout. An obvious consequence of the noisy hardware is a non-vanishing probability of leakage out of the physical Hilbert space, as the associated errors can violate gauge invariance. One can, therefore, consider a ``symmetry-based'' error mitigation scheme, which is to only count the events within the physical Hilbert space, and properly renormalize the final wave-function outcome. For wave packets in panels (a) and (b) of Fig.~\ref{fig:wp_all}, respectively, 356 and 412 events out of the 500 shots are left within the physical Hilbert space after such an error-mitigation protocol, and 306 and 349 events have their ancilla measured to be 1.  

Due to resource limitations, we estimate the uncertainty associated with the hardware results by a bootstrap procedure for each wave-packet simulation. Specifically, to exclude the error that is definitely hardware-systematic, only the physical events are bootstrapped. In each physical bootstrap sample, the events with the ancilla measured to be 0 (due to the residual errors) are excluded, and the remaining probabilities are normalized and collected. For both wave packets, $10^4$ resamplings are used to ensure bootstrap-sample mean distributions, hence the standard deviations, are stabilized. In Fig.~\ref{fig:wp_all}, the uncertainties on the probabilities are the standard deviation of the bootstrap resampling, on which a standard error propagation gives the uncertainties on the staggered density. 

The physical basis-states probabilities show acceptable agreement with the truncated-circuit results obtained via statevector evolution. Perhaps a more meaningful comparison is with the result obtained from a classical circuit simulator that uses the same number of measurement shots as that in the hardware implementation. Such `noiseless` simulation results are presented in Appendix~\ref{app:noisy_emu}. We further employ the Quantinuum's emulator to inspect how accurately it agrees with the hardware results for the circuits implemented in this work, and present the result in the same Appendix. In both cases, the uncertainty estimation described above using bootstrap resampling is applied.

Currently, we could only afford to create a single wave packet for two sets of parameters on Quantinuum's quantum computer. Performing more sophisticated measurement schemes, such as energy-density measurements and other non-local correlators, as well as full entanglement tomography of the resulting wave packets, would require more computational resources that can be carried out in future work.

\section{Summary and outlook}
\label{sec:summary}
\noindent
We demonstrated how interacting hadronic wave packets within confined lattice gauge theories in 1+1 dimensions can be prepared on a quantum computer without adiabatic evolution. This is achieved by directly building the interacting mesonic excitations via an optimized ansatz. Our state preparation involves generation of the interacting ground state using a variational quantum eigensolver, optimization of a mesonic excitation ansatz in each momentum sector in the interacting theory, then building and circuitizing the full Gaussian wave-packet creation operator with given width in position (or momentum) space using efficient algorithms. We choose a $Z_2$ LGT in 1+1 D with one flavor of staggered fermions to test our method for its simplicity, nonetheless, we also constructed and tested numerically the ansatz for the case of the (single-flavor) Schwinger model. Exact optimized wave packets, as well as those obtained from circuits with both infinite-shot and finite-shot statistics are created on a 6-site lattice. The quantum circuits encode the wave-packet creation operator using a single ancilla qubit, and the number of entangling gates is minimized by a singular-value decomposition algorithm. The interacting mesonic creation operator involves many bare mesonic operators. Upon prioritizing mesons with smaller sizes, the wave-packet circuits were executed on the Quantinuum \texttt{H1-1} system. After excluding the probabilities outside of the physical Hilbert space, the hardware result is found to be in agreement with classical simulation.

The algorithm of this work, therefore, provides an efficient and effective procedure to create hadronic wave-packet states for scattering in (1+1)D gauge theories. While ground and low-energy state preparation in such models is feasible with classical methods, such as tensor networks,  of particular interest are scattering processes at high energies with abundant entanglement generation, which are challenging for classical algorithms. Hence, combining an efficient state-preparation algorithm, such as that provided in this work, and time-evolution and measurement steps on a quantum computer, can facilitate studies of scattering on quantum hardware in the future.

This work can be expanded in several directions in future studies:
\begin{itemize}
\item[$\diamond$]{
We focused on simple local observables, such as basis-states probabilities, energy density, and local staggered charge density, to assess the quality of the wave packet generated on the hardware. In principle, other two- and $n$-point functions, including the non-local ones, need to be measured to fully verify the fidelity of the state produced. These, in turn, require additional quantum-circuit runs involving measurement of the desired operators, which were too costly given our limited computational credits on Quantinuum. Alternatively, one may resort to efficient entanglement-Hamiltonian tomography schemes~\cite{Kokail:2020opl,Kokail:2021ayb,Joshi:2023rvd,Mueller:2022xbg,Bringewatt:2023xxc}, augmented by randomized measurement tools~\cite{Ohliger_2013,Elben:2019dxb,Huang:2020tih,Vermersch:2023gki,Elben:2022jvo}, to reconstruct the system's density matrix. These require many random measurements, and deemed expensive in our current study using Quantinuum's resources. Such comprehensive measurement schemes, nonetheless, will be essential for the full scattering process, as deducing the nature of the post-collision state is an ultimate goal of the simulation.}
\item[$\diamond$]{State-of-the-art quantum hardware may soon reach capacity and capability needed to host larger systems. Therefore, wave-packet preparation in a confined gauge theory with a continuum limit, such as the $U(1)$, $SU(2)$, and $SU(3)$ LGTs in 1+1 D, should be feasible [see authors' note below]. There is no qualitative difference in terms of circuit design, except bosonic operations are performed on a multi-qubit subspace for each gauge field. Various subcircuits for relevant operations, such as boson addition and subtraction, have, nonetheless, been developed in literature~\cite{Shaw:2020udc,Davoudi:2022xmb} and can be straightforwardly ported to the algorithm of this work. Importantly, generating mesonic excitations in higher dimensional theories can likely proceed via the ansatz of this work, as mesons should still be well approximated by quark-antiquark pairs connected by an electric flux. Nonetheless, there will be more bare mesonic operators to account for different shapes and orientations of the fluxes in space.}
\item[$\diamond$]{While we have carefully assessed the algorithmic and statistical errors associated with our construct in the numerical benchmarks performed on small systems, it would be favorable to conduct a comprehensive analytical study of systematic errors introduced by the inexact exponentiation and Trotterization, see Sec.~\ref{sec:3-1}. Such studies are common in the context of implementing the time-evolution operator within given accuracy (see e.g., Refs.~\cite{Shaw:2020udc,Kan:2021xfc,Davoudi:2022xmb} for gauge-theory examples) and need to be extended to state-preparation studies in gauge theories. Only such studies will enable accurate estimates of the resources required as a function of the desired wave-packet quality, and will allow comparing the efficiency of different methods in literature in near- and far-term eras of quantum computing. We defer such an analysis to future work.
}
\item[$\diamond$]{A natural next step of our algorithm is to prepare two wave-packet states, followed by time evolution to realize scattering on quantum computers. Once the hadronic creation operator coefficients $C_{m,n}$ are determined, any number of wave packets can be created along the lattice, as long as they are well separated initially. One can thus reconnect to the Jordan-Lee-Preskill protocol for the S-matrix construction, with the state preparation bottleneck now largely reduced. To create a meaningful scattering experiment, one would need wave packets with larger extents in position space (e.g. 10 staggered sites) to ensure a narrower width in momentum space. Creating two 10-site wave packets with sufficiently large separation would require hundreds of qubits and thousands of entangling gates (assuming the circuits are truncated to 1-length bare mesons). Together with the subsequent time evolution and measurements, the total resources may exceed the current capacity of quantum hardware, but could soon be within reach. We note that for high-energy scattering, mesons with large values of momenta need to be created, which may prove challenging to produce with high fidelity using the ansatz of this work.}
\item[$\diamond$]{While we did not discuss analog-simulation schemes in this work, we should remark that several interesting proposals have been developed in recent years to demonstrate analog simulators as a useful tool for studying scattering processes in simple gauge theories. For example, Ref.~\cite{Surace:2020ycc} presents protocols to observe and measure selected meson-meson scattering processes in a $Z_2$ model in 1+1 D in Rydberg-atom arrays. Reference~\cite{Belyansky:2023rgh} proposes a scheme to simulate collision of high-energy quark-antiquark and meson-meson wave packets in non-confining and confining regimes of the lattice Schwinger model with a topological $\theta$-term using circuit-QED platforms, after mapping the model to a bosonic equivalent. Furthermore, Ref.~\cite{Su:2024uuc} proposes a particle-collision experiment in optical-superlattice quantum simulators of a lattice Schwinger model with a $\theta$-term. It would be interesting to see how the mesonic ansatz of this work can facilitate these and future analog and hybrid analog-digital simulations of scattering processes in various LGTs.
}
\item[$\diamond$]{The ultimate goal is to develop expressive and efficient forms of hadronic wave-packet ansatze, particularly for baryons and nuclei, relevant to quantum chromodynamics. Insights from classical LGT studies of hadrons and nuclei will likely be crucial in coming up with such ansatze. In particular, a wealth of results on the spectrum~\cite{Detmold:2019ghl,Bulava:2022ovd,Prelovsek:2023sta} and structure~\cite{Constantinou:2020hdm,Hagler:2009ni} of hadrons and nuclei using lattice-QCD computations may need to be incorporated in state-preparation schemes on quantum computers to reduce the required quantum resources. These results can potentially both inform the form of the hadronic ansatz, and aid in parameter optimizations. The strategy of this work can be generalized to take advantage of such opportunities.
}
\end{itemize}

\vspace{0.5 cm}
\noindent
\emph{Authors' note.} In the final stage of drafting this manuscript, another manuscript on wave-packet preparation in the Schwinger model was released~\cite{Farrell:2024fit}. This reference prepares the wave packet in the interacting theory using an improved VQE algorithm that finds a low-depth quantum circuit which creates the wave packets out of the interacting vacuum. The interacting vacuum itself is prepared using the improved VQE developed by the authors. The optimization for the VQE is performed by comparing against a wave packet that is adiabatically produced out of a bare mesonic wave packet with the aid of classical computing. Using such a hybrid classical-quantum strategy, and performing state-of-the-art noise mitigation techniques suited to the IBM quantum-computing platform, the authors create and evolve single wave packets on lattices as large as 112 staggered sites (112 qubits). We leave comparing the quality of wave packets prepared via this strategy and that presented in our work to future studies.

\section*{Acknowledgment}
\noindent
We thank Brian Neyenhuis and Jenni Strabley from Quantinuum for their generosity and support, and Microsoft Azure and the Division of Information Technology at the University of Maryland (UMD IT) for providing the infrastructures to access the quantum emulator and hardware via Cloud Services.
We are grateful to Niklas Mueller for his early effort in ensuring access to the Azure Cloud subscription via UMD IT, and for the help in acquisition of some of the quantum-computing credits used in this project. We further thank Connor Powers for assistance with aspects of running our codes on a local computing machine and using Azure Cloud Services. We further acknowledge useful discussions with Navya Gupta, Jesse Stryker, and Francesco Turro. The numerical calculations presented in this work are made possible by the extensive use of \texttt{Qiskit}~\cite{Qiskit} and \texttt{Jupyter Notebook}~\cite{kluyver2016jupyter} software applications within the \texttt{Conda}~\cite{conda} environment.
Z.D., C-C.H., and S.K. acknowledge funding by the U.S. Department of Energy (DOE), Office of Science, Early Career Award (award no. DESC0020271), and the DOE, Office of Science, Office of Nuclear Physics, via the program on Quantum
Horizons: QIS Research and Innovation for Nuclear Science (award no. DE-SC0021143). Z.D. and C-C.H. are grateful for the hospitality of Perimeter Institute where part of this work was carried out. Research
at Perimeter Institute is supported in part by the Government of Canada through the Department of
Innovation, Science, and Economic Development, and by the Province of Ontario through the Ministry of Colleges and Universities. This research was also supported in part by the Simons Foundation through the Simons Foundation Emmy Noether Fellows Program at Perimeter Institute. S.K. further acknowledges support by the U.S. DOE, Office of Science, Office of Nuclear Physics, InQubator for Quantum Simulation (IQuS) (award no. DE-SC0020970), and by the DOE QuantISED program through the theory consortium ``Intersections of QIS and Theoretical Particle Physics'' at Fermilab (Fermilab subcontract no. 666484).
This work was also supported, in part, through the Department of Physics and the College of Arts and Sciences at the University of Washington.

\bibliography{bibi.bib}


\newpage
\appendix
\section{Numerical verification of the ansatz for $U(1)$ LGT}
\label{app:u1-ansatz}
\begin{figure}[h!]
    \centering
    \includegraphics[width=1\textwidth]{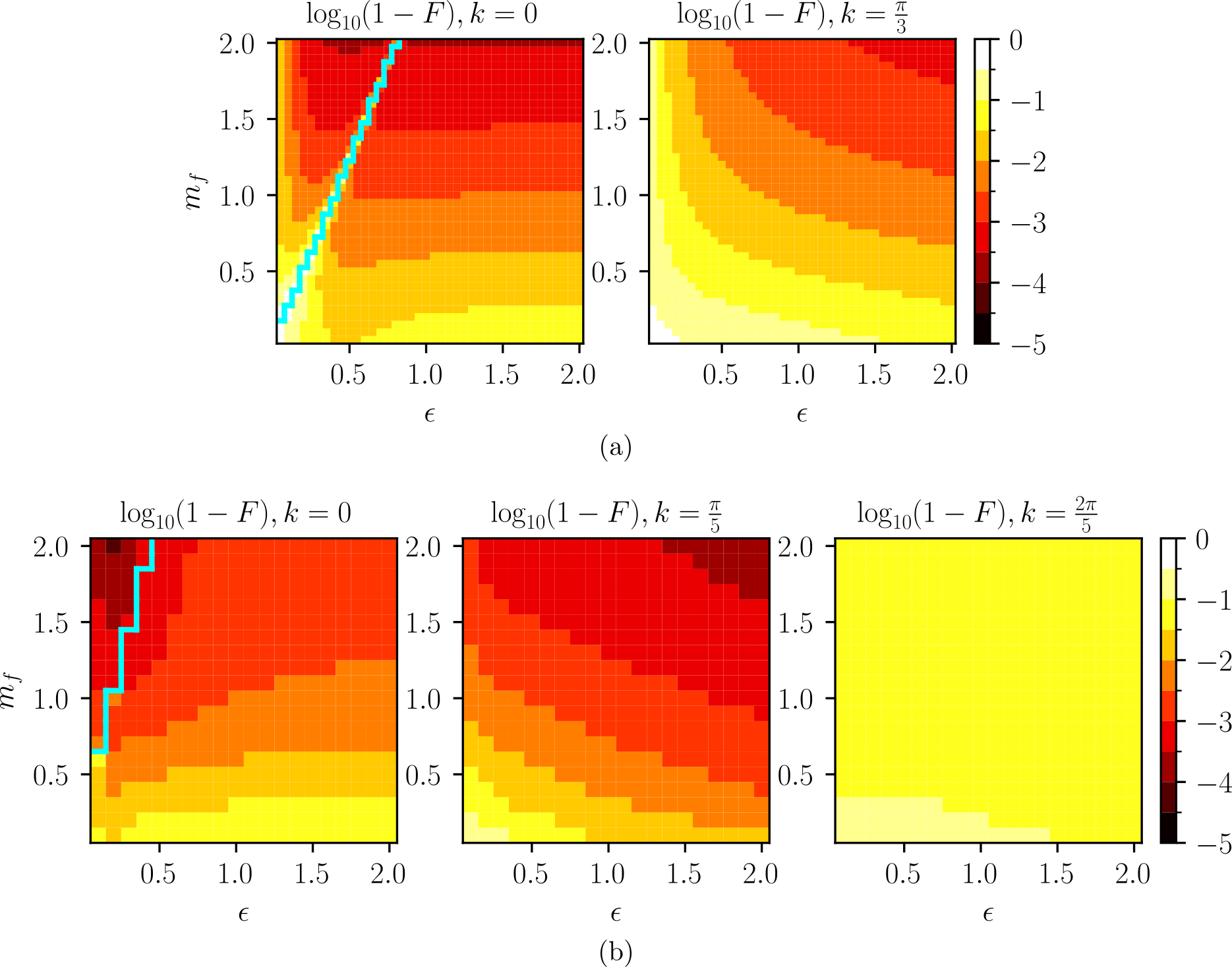}
    \caption{The $k$-optimization heat map for the $U(1)$ LGT. Panels (a) and (b) refer to the 6-site and the 10-site theories, respectively (with a cutoff of $\Lambda=1$). The region left to the cyan contour is where the purely non-mesonic states are less energetic than the single-meson $\ket{k}$ state. The optimization works well for the 6-site theory and $\ket{k=0}$ and $\ket{k=\pm\frac{\pi}{5}}$ in the 10-site theory, but is less effective for $\ket{k=\pm\frac{2\pi}{5}}$, where fidelities of only $90\%$ are reached for most of the parameter space.}
    \label{fig:scan_10s}
\end{figure}

\noindent
Supplementing the results of Sec.~\ref{sec:2-3}, in this Appendix we provide the ansatz-optimization results for the $U(1)$ LGT in 1+1 D. We consider the 6-site and the 10-site theories using a cutoff of $\Lambda=1$. Figure~\ref{fig:scan_10s} shows the fidelity between the optimized $\ket{k}$ and the exact-diagonalized momentum eigenstates for the same Brillouin zone as in Fig.~\ref{fig:scan}. Results for negative $k$ values are identical to those for the respective positive values and are not shown.

\begin{figure}
    \centering
    \includegraphics[scale=0.95]{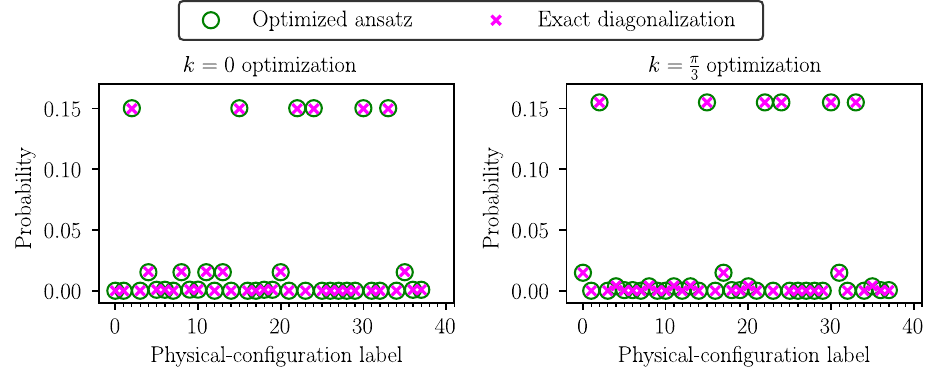}
    \caption{Optimization results for $k=0~(\frac{\pi}{3})$ momentum eigenstates are shown in the left (right) panel for the case of 6-site $U(1)$ LGT with $m_f=1$, $\epsilon=1.0$, and cutoff $\Lambda=1$. The physical basis-state probabilities the of $k=-\frac{\pi}{3}$ state are identical to $k=\frac{\pi}{3}$ and are not shown. The physical basis states of the 6-site $U(1)$ theory are listed in Tables~\ref{tab:physical_u1}.}
    \label{fig:k_optimization_u1}
\end{figure}

Similar to Fig.~\ref{fig:scan} for the $Z_2$ LGT, the region where the low-lying purely non-mesonic states are present is marked to the left side of the cyan line. The energy required to excite such states from the interacting ground state is proportional to the system size. We observed the same reduction of the non-mesonic excitation regime from 6 sites to 10 sites as in the case of $Z_2$ LGT. Finally, we plot in Fig.~\ref{fig:k_optimization_u1} the probability distribution of each of the physical configurations in the $U(1)$ theory with parameters $m_f=1$ and $\epsilon=1.0$. Each $k$-momentum state created by the ansatz recovers the exact-diagonalized state probabilities.

%

\section{Efficient-circuit parameters}
\label{app:truncation}
\noindent
We discuss the effect of circuit parameters $N_{\text{Trotter}}$ and $\theta_c$ in this Appendix for the case of the $Z_2$ LGT. The former introduces a Trotter error when implementing $e^{-i\frac{\pi}{2}\Theta}=e^{-i\frac{\pi}{2}\sum_{m,n}\Theta_{m,n}}$ in Eq.~\eqref{eq:anc1} via a second-order Trotter formula. The latter truncates bare mesons with small coefficient $C_{m,n}$ in $\Theta_{m,n}$, see Eq.~\eqref{eq:Theta-m-n}. We use the wave packet with parameters $\sigma=\frac{\pi}{6}$, $\mu=3$, and $k_0=0$ as an example.

\begin{figure}[hbt!]
    \centering
    \includegraphics[]{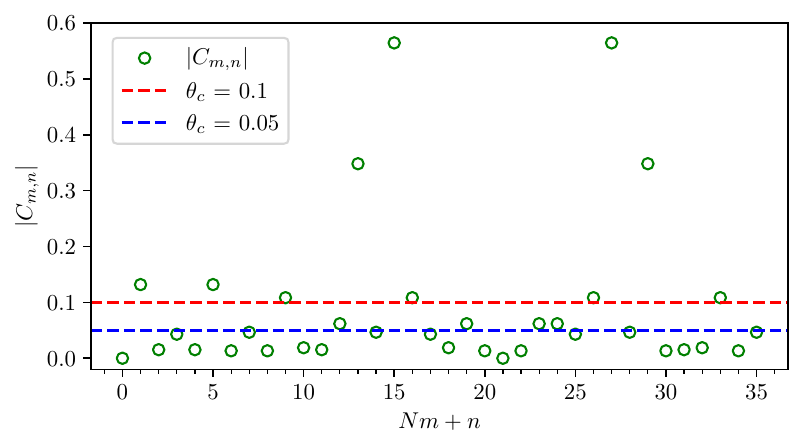}
    \caption{The magnitudes of all $C_{m,n}$ coefficients in the case of the 6-site $Z_2$ LGT, which translate to the rotation angles in the wave-packet circuit [see Eq.~\eqref{eq:svd_D}]. The x-axis represents a total of $N^2$ position pairs  $(m,n)$, i.e., 36 for 6 sites. The horizontal lines indicate different $\theta_c$ values, which if imposed on the circuits, only the mesons associated with the $|C_{m,n}|$ values above them are built. The hardware simulations of this work use $\theta_c=0.1$ to keep the circuits sufficiently shallow.}
    \label{fig:angle}
\end{figure}

Figure~\ref{fig:angle} shows $|C_{m,n}|$, the magnitude of the amplitude of the bare mesons generated by $\widetilde{\mathcal{M}}_{m,n}$ in position space. Only a few out of all $N^2$ possible mesons dominate the wave packet, and these are primarily 1-length mesons. Thus, the wave-packet circuit can be efficiently implemented using a non-zero $\theta_c$. Note that according to Table~\ref{tabel:gate-count}, implementing the full meson wave packets amounts to implementing $\mathcal{O}(N^3)$ entangling gates, while accounting for only 1-length mesons in the ansatz requires $\mathcal{O}(N)$ entangling gates, at the cost of reduced accuracy, as demonstrated in the numerical study of this work.

\begin{figure}
    \centering
\includegraphics[width=1\textwidth]{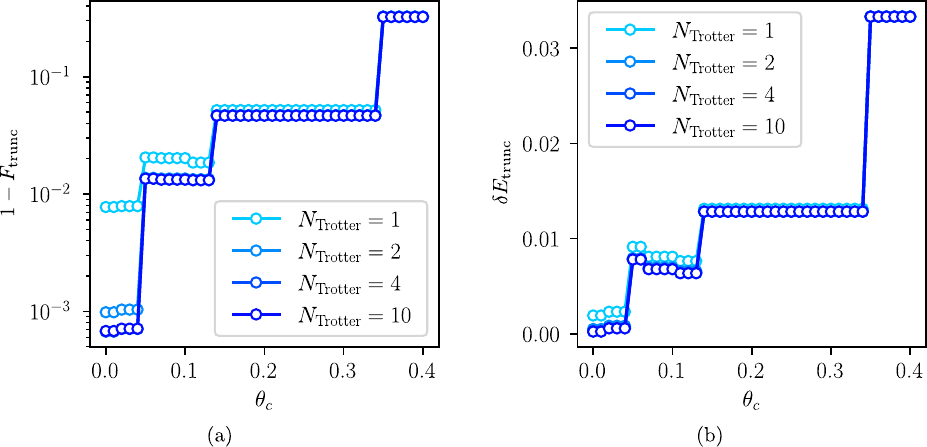}
    \caption{The dependence of fidelity and relative energy difference on $N_{\text{Trotter}}$ and $\theta_c$ are presented in panels (a) and (b), respectively. The improvement from more $N_{\text{Trotter}}$ using the second-order product formula is immediate: 2 Trotter steps are nearly enough to converge to the ideal result. (Note that in panel (a), the values for $N_\text{Trotter}=4$ and $N_\text{Trotter}=10$ are overlapping. For panel (b), all values except for those for $N_\text{Trotter}=1$ overlap). Reducing $\theta_c$ sequentially accepts more mesons in groups, resulting in a step-wise improvement.}
    \label{fig:trunc_f_e}
\end{figure}

The effect of different $N_{\text{Trotter}}$ and $\theta_c$ on fidelity and relative energy difference is shown in Fig.~\ref{fig:trunc_f_e}. Here, the relative energy difference is defined as
\begin{align}
\delta E_{\rm trunc} = \frac{|E_{\text{trunc}}-E_{\text{exact}}|}{|E_{\text{exact}}|},
\label{eq:energy_app}
\end{align}
where $E_\text{exact}$ ($E_\text{trunc}$) is the Hamiltonian expectation value using $\ket{\Psi}_\text{exact}$ ($\ket{\Psi}_\text{trunc}$), while the fidelity is defined as
\begin{align}
F_{\rm trunc}=\left|{}_\text{trunc}\langle \Psi | \Psi \rangle_\text{exact}\right|^2.
\label{eq:fidelity_app}
\end{align}
The results for both quantities saturate quickly as $N_{\text{Trotter}}$ increases. As $\theta_c \rightarrow 0$, the quality of states improves step-wise due to the gaps in the $|C_{m,n}|$ values as is seen in Fig.~\ref{fig:angle}.

\begin{figure}
    \centering
\includegraphics[scale=1.0]{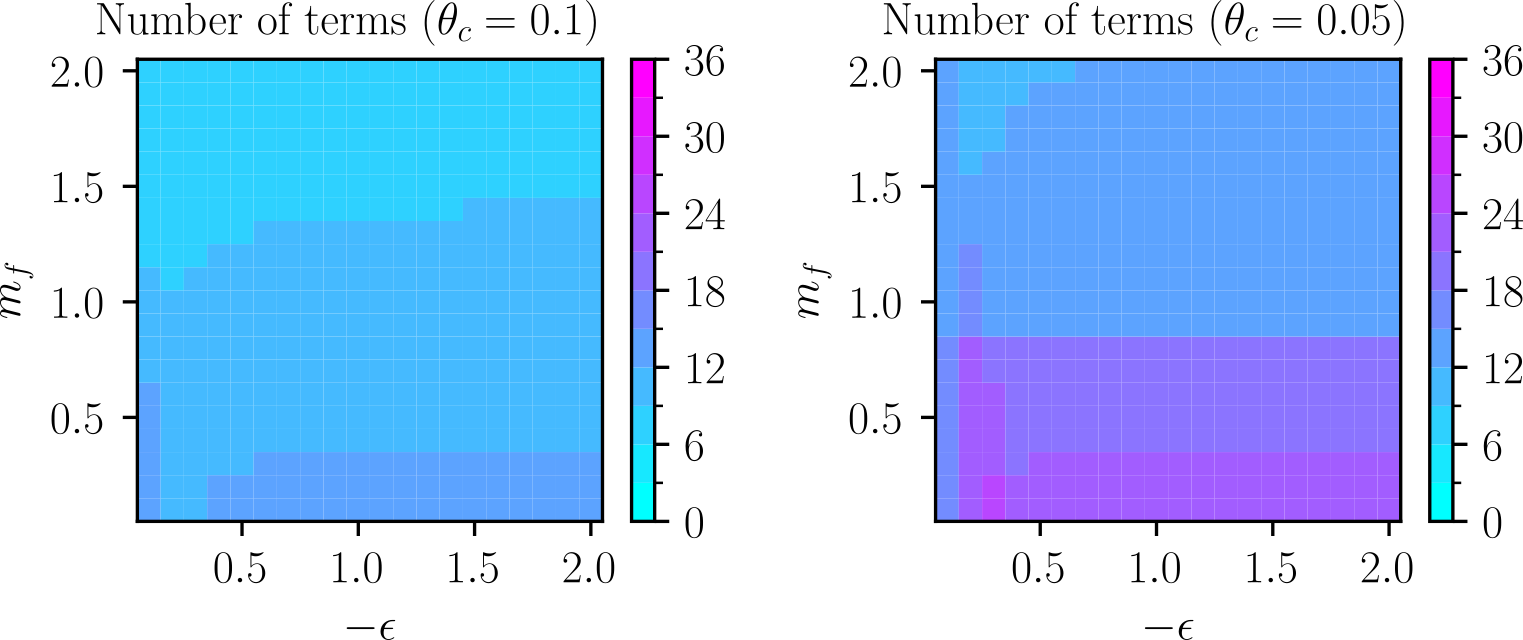}
    \caption{The number of mesonic creation operators using the wave-packet parameters ($\sigma=\frac{\pi}{6}, \mu=3, k_0=0$) but with a wide range of $m_f$, $\epsilon$ values is plotted. The left (right) panels corresponds to $\theta_c=0.1(0.05)$. The maximum number of terms is $N^2$ (36 for the 6-site theory). It is observed that more terms survive the $\theta_c$ truncation in the low-mass regime.}
    \label{fig:scan_cmn}
\end{figure}

The analysis above considers the theory with $m=1.0$, $\epsilon=-0.3$. Additionally, the effect of $\theta_{c}$ truncation for the same wave-packet parameters but different $m_f$ and $\epsilon$ values is shown in Fig.~\ref{fig:scan_cmn}. Qualitatively, there are more terms surviving the truncation in the low-mass regime.

%
\section{Finite measurement-shot analysis
\label{app:shots}}
\noindent
In this appendix, we provide justification of the number of measurement shots used for hardware experiments of this work. The wave-packet basis-state probabilities obtained from the hardware suffers from two sources of error against the exact statevector-evolution result. The statistical error induced by the finite number of shots, and the systematic error from the hardware noise. The former can be reduced by increasing the number of shots, while the latter can only be improved by eliminating the hardware error. The simple error-mitigation scheme considered in this work (projection to the physical Hilbert space by post-processing) turned out to be sufficient to provide reasonable agreement with theoretical values, so here we focus on the effect of the statistical shot noise.

The fidelity used to describe the quality of the wave packets previously is inapplicable to the finite-shot results without any state tomography, as the phases are lacking from the probability distributions. Instead, we consider the root-mean-square (rms) error as an alternative measure for the discrepancy in probabilities:
\begin{equation}
    \varepsilon_{\rm rms} = \sqrt{\sum_{i=0}^{N_{\mathcal{H}}}{\frac{{(P^i_{\rm shots}-P^i_{\rm trunc})^2}}{N_{\mathcal{H}}}}}.
    \label{eq:epsilon-rms}
\end{equation}
Here, $N_{\mathcal{H}}$ is the size of the physical Hilbert space, $P^i_{\rm shots(trunc)}$ represents the overlap probability of the finite-shot (truncated-circuit) results onto the $i^{\rm th}$ configuration in the physical Hilbert space. The truncated circuit is seen as the ``ground truth'' in this expression, as it is the one implemented on the hardware.

We expect $\varepsilon_{\rm rms}$ to decrease as $N_{\rm shots}$ increases, and saturate at some number $N_{\rm shots}^*$, where the statistical error becomes less significant than the residual systematic hardware error. The truncated circuit is run with different $N_{\rm shots}$ values using the Quantinuum \texttt{H1-1} emulator. $N_{\rm shots}^*$, once determined, should set the number of shots used in the experiment, as there is no point of expanding the sample size beyond this value. Furthermore, one can calculate the rms error between the truncated-circuit and the exact-optimized states' probabilities by replacing $P^i_{\rm trunc}$ with $P^i_{\rm exact}$ in Eq.~\eqref{eq:epsilon-rms}, labeled $\varepsilon_{\rm rms}^*$. Similar to the fidelity calculated in the main text, $\varepsilon_{\rm rms}^*$ characterizes the systematic error due to the truncation, and can be used as another reference for choosing $N_{\rm shots}$. The finite-shot result is seen to be good enough if it differs from the statevector evolution by an amount comparable to $\varepsilon_{\rm rms}^*$, the difference between the latter and the exact-optimized wave packet.

\begin{figure}
    \centering
    \includegraphics{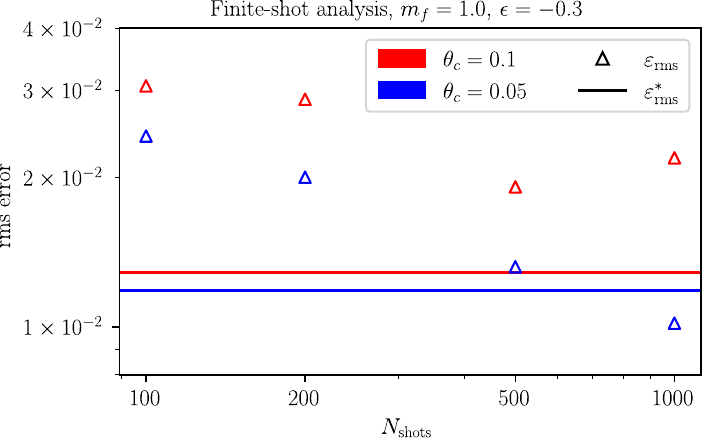}
    \caption{The rms errors as described in the text as a function of the number of measurement shots. The values of $N^*_{\rm shots}$ to be taken for hardware experiments are where $\varepsilon_{\rm rms}$ saturates or touches $\varepsilon_{\rm rms}^*$, which ever comes first. For $\theta_c=0.1$ (red), the first condition is met at $N^*_{\rm shots}=500$. For $\theta_c=0.05$ (blue), the second condition is met at $N^*_{\rm shots}=1000$,  before the error saturates.}
    \label{fig:rms}
\end{figure}

In Fig.~\ref{fig:rms}, the rms error obtained from the circuit with truncation $\theta_c=0.1$ (red, the one used in this work), indeed stops decreasing above $N_{\rm shots}^*=500$. As a result, we choose the number of shots for hardware experiments to be 500. The horizontal line represents $\varepsilon_{\rm rms}^*$, which is slightly below $\varepsilon_{\rm rms}$ for $\theta_c=0.1$, the least truncation we can afford with current resources. For comparison, emulator results using $\theta_c=0.05$, labeled with blue, are also included. Although $\varepsilon_{\rm rms}$ keeps improving, it reaches $\varepsilon_{\rm rms}^*$ at $N_{\rm shots}^*=1000$. Therefore, to perform an experiment with this less-truncated, hence more accurate, circuit in the future, the recommended number of shots becomes around 1000.
%

\section{Wave packets with quantum-hardware emulators}
\label{app:noisy_emu}
\noindent
\begin{figure}[t!]
    \centering
    \includegraphics[width=1\textwidth]{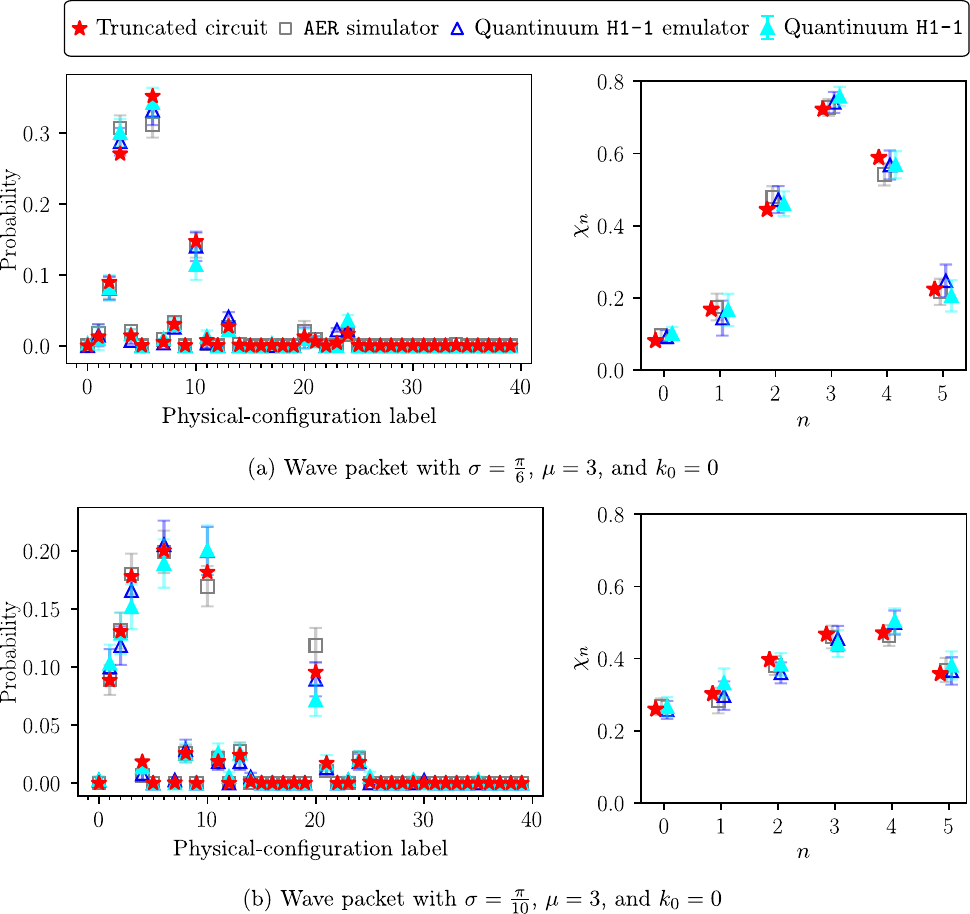}
    \caption{
    Physical basis-state probabilities of wave packets with two sets of parameters generated on the Quantinuum \texttt{H1-1} quantum emulator and computer compared with those obtained from \texttt{AER} simulator and from $\ket{\Psi}_\text{trunc}$, as well as the associated local particle densities, $\chi_n$. All data except for those from $\ket{\Psi}_\text{trunc}$ correspond to 500 measurement shots. The Quantinuum emulator and computer results are obtained upon a symmetry-based error mitigation as discussed in the main text. The physical basis states are listed in Table~\ref{tab:physical_z2} in Appendix~\ref{app:state-labels}.
    }
    \label{fig:wp_noisy_and_hardware}
\end{figure}
In this appendix, we investigate the effect of statistical shot noise compared with the hardware-specific sources of error. To this aim, we present the same quantities as in Fig.~\ref{fig:wp_all} of the main text in Fig.~\ref{fig:wp_noisy_and_hardware}, removing the exact and ideal-circuit results, while adding Qiskit's \texttt{AER} simulator~\cite{Qiskit} with 500 measurement shots (as used to obtain the hardware results). The \texttt{AER} simulator provides values affected by only the statistical shot noise. It is observed that the effect of shot noise is generally small, and the hardware-specific error likely contributes more to the disagreement between the statevector-evolution results and the hardware results.

\begin{figure}[t!]
    \centering
    \includegraphics[width=1\textwidth]{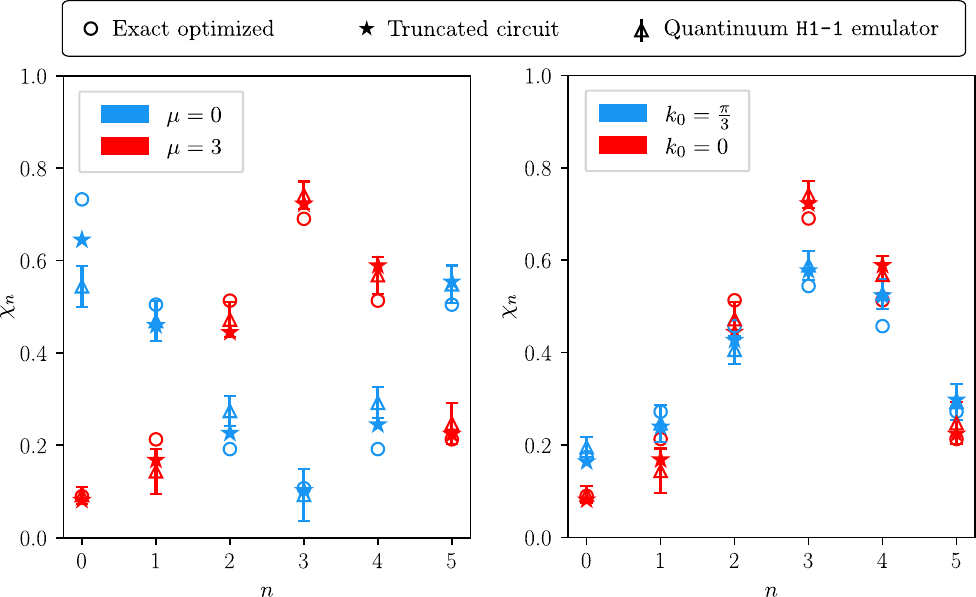}
    \caption{
    Local particle density $\chi_n$ for wave packets with varying $\mu$ and $k_0$, obtained from Quantinuum \texttt{H1-1} emulator compared with those obtained from the optimized wave-packet state and the 1-meson truncated-circuit output. For both plots, $\sigma = \frac{\pi}{6}$, while for the left plot, $k_0=0$ and and for the right plot, $\mu=3$.
    }
    \label{fig:diff_wp}
\end{figure}

It is also valuable to compare the values obtained by the Quantinuum \texttt{H1-1} emulator, which incorporates both the statistical shot noise and hardware-specific noise. The latter is modeled to faithfully emulate the hardware performance. These results are also included in the plots in Fig.~\ref{fig:wp_noisy_and_hardware}. There is a reasonable agreement between the emulator and the hardware results, indicating that the noisy emulator describes the quantum hardware faithfully. The uncertainties on the \texttt{AER} simulator and the Quantinuum \texttt{H1-1} emulator results are obtained using the same bootstrap procedure described for the hardware results in Sec.~\ref{sec:4-2}. Since we have limited hardware access but less restriction on emulator access, we can perform further studies with this emulator. As an example, to verify the behavior of the wave-packet preparation circuit as a function of other parameters of the wave packet beside its width, we have obtained the local fermionic density $\chi_n$, defined in Eq.~\eqref{eq:chi1}, for different values of $\mu$ and $k_0$. The results are plotted in Fig.~\ref{fig:diff_wp}. 500 shots are used for each parameter set $(\sigma, \mu, k_0)$, and the physical error mitigation mentioned in Sec.~\ref{sec:4-2} is applied to the Quantinuum \texttt{H1-1} emulator results. The results are in agreement with the expectations: changing $\mu$ corresponds to changing the center of the wave packet in position space, while changing $k_0$ corresponds to changing the wavenumber in position space.

%

\section{Physical Hilbert space of the 6-site theories
\label{app:state-labels}}
\noindent
In this appendix, we list all 40 (38) physical configurations that satisfy the Gauss' law [Eq.~\eqref{eq:gauss}] in the 6-site $Z_2$ LGT ($U(1)$ LGT) in 1+1 D with one flavor of staggered fermions. The assigned labels of these configurations (i.e., physical basis states) are used as the x-axis in Figs.~\ref{fig:k_optimization}, \ref{fig:GS_prob}, \ref{fig:wp_all}, and \ref{fig:wp_noisy_and_hardware}. As mentioned in the main text, the physical Hilbert space is restricted to the subspace with $Q=N/2$, where the fermion-number operator $Q$ is defined in Sec.~\ref{sec:2-1}. For the $U(1)$ LGT, the electric-field basis is cut off at the value $\Lambda=1$, allowing only three possible electric-field values ($0,\pm 1$) at each link.
\begin{table}[ht!]
    \centering
    \begin{tabular}{|c||*{12}{c|}}\hline
Physical- & \multirow{3}{*}{$\ket{0}_{f}$} & \multirow{3}{*}{$\ket{0}_{b}$} & \multirow{3}{*}{$\ket{1}_{f}$} & \multirow{3}{*}{$\ket{1}_{b}$} & \multirow{3}{*}{$\ket{2}_{f}$} & \multirow{3}{*}{$\ket{2}_{b}$} & \multirow{3}{*}{$\ket{3}_{f}$} & \multirow{3}{*}{$\ket{3}_{b}$} & \multirow{3}{*}{$\ket{4}_{f}$} & \multirow{3}{*}{$\ket{4}_{b}$} & \multirow{3}{*}{$\ket{5}_{f}$} & \multirow{3}{*}{$\ket{5}_{b}$} \\
configura- & & & & & & & & & & & & \\
tion label & & & & & & & & & & & & \\
\hline
0 & 0 & 0 & 1 & 0 & 0 & 0 & 1 & 0 & 0 & 0 & 1 & 0 \\ \hline
1 & 1 & 1 & 0 & 0 & 0 & 0 & 1 & 0 & 0 & 0 & 1 & 0 \\ \hline
2 & 0 & 0 & 0 & 1 & 1 & 0 & 1 & 0 & 0 & 0 & 1 & 0 \\ \hline
3 & 0 & 0 & 1 & 0 & 1 & 1 & 0 & 0 & 0 & 0 & 1 & 0 \\ \hline
4 & 1 & 1 & 0 & 0 & 1 & 1 & 0 & 0 & 0 & 0 & 1 & 0 \\ \hline
5 & 1 & 1 & 1 & 1 & 0 & 1 & 0 & 0 & 0 & 0 & 1 & 0 \\ \hline
6 & 0 & 0 & 1 & 0 & 0 & 0 & 0 & 1 & 1 & 0 & 1 & 0 \\ \hline
7 & 1 & 1 & 0 & 0 & 0 & 0 & 0 & 1 & 1 & 0 & 1 & 0 \\ \hline
8 & 0 & 0 & 0 & 1 & 1 & 0 & 0 & 1 & 1 & 0 & 1 & 0 \\ \hline
9 & 0 & 0 & 0 & 1 & 0 & 1 & 1 & 1 & 1 & 0 & 1 & 0 \\ \hline
10 & 0 & 0 & 1 & 0 & 0 & 0 & 1 & 0 & 1 & 1 & 0 & 0 \\ \hline
11 & 1 & 1 & 0 & 0 & 0 & 0 & 1 & 0 & 1 & 1 & 0 & 0 \\ \hline
12 & 0 & 0 & 0 & 1 & 1 & 0 & 1 & 0 & 1 & 1 & 0 & 0 \\ \hline
13 & 0 & 0 & 1 & 0 & 1 & 1 & 0 & 0 & 1 & 1 & 0 & 0 \\ \hline
14 & 1 & 1 & 0 & 0 & 1 & 1 & 0 & 0 & 1 & 1 & 0 & 0 \\ \hline
15 & 1 & 1 & 1 & 1 & 0 & 1 & 0 & 0 & 1 & 1 & 0 & 0 \\ \hline
16 & 1 & 1 & 1 & 1 & 1 & 0 & 0 & 1 & 0 & 1 & 0 & 0 \\ \hline
17 & 0 & 0 & 1 & 0 & 1 & 1 & 1 & 1 & 0 & 1 & 0 & 0 \\ \hline
18 & 1 & 1 & 0 & 0 & 1 & 1 & 1 & 1 & 0 & 1 & 0 & 0 \\ \hline
19 & 1 & 1 & 1 & 1 & 0 & 1 & 1 & 1 & 0 & 1 & 0 & 0 \\ \hline
20 & 1 & 0 & 1 & 0 & 0 & 0 & 1 & 0 & 0 & 0 & 0 & 1 \\ \hline
21 & 1 & 0 & 0 & 1 & 1 & 0 & 1 & 0 & 0 & 0 & 0 & 1 \\ \hline
22 & 0 & 1 & 1 & 1 & 1 & 0 & 1 & 0 & 0 & 0 & 0 & 1 \\ \hline
23 & 1 & 0 & 1 & 0 & 1 & 1 & 0 & 0 & 0 & 0 & 0 & 1 \\ \hline
24 & 1 & 0 & 1 & 0 & 0 & 0 & 0 & 1 & 1 & 0 & 0 & 1 \\ \hline
25 & 1 & 0 & 0 & 1 & 1 & 0 & 0 & 1 & 1 & 0 & 0 & 1 \\ \hline
26 & 0 & 1 & 1 & 1 & 1 & 0 & 0 & 1 & 1 & 0 & 0 & 1 \\ \hline
27 & 0 & 1 & 0 & 0 & 1 & 1 & 1 & 1 & 1 & 0 & 0 & 1 \\ \hline
28 & 1 & 0 & 0 & 1 & 0 & 1 & 1 & 1 & 1 & 0 & 0 & 1 \\ \hline
29 & 0 & 1 & 1 & 1 & 0 & 1 & 1 & 1 & 1 & 0 & 0 & 1 \\ \hline
30 & 0 & 1 & 0 & 0 & 0 & 0 & 1 & 0 & 1 & 1 & 1 & 1 \\ \hline
31 & 0 & 1 & 0 & 0 & 1 & 1 & 0 & 0 & 1 & 1 & 1 & 1 \\ \hline
32 & 1 & 0 & 0 & 1 & 0 & 1 & 0 & 0 & 1 & 1 & 1 & 1 \\ \hline
33 & 0 & 1 & 1 & 1 & 0 & 1 & 0 & 0 & 1 & 1 & 1 & 1 \\ \hline
34 & 1 & 0 & 1 & 0 & 0 & 0 & 0 & 1 & 0 & 1 & 1 & 1 \\ \hline
35 & 1 & 0 & 0 & 1 & 1 & 0 & 0 & 1 & 0 & 1 & 1 & 1 \\ \hline
36 & 0 & 1 & 1 & 1 & 1 & 0 & 0 & 1 & 0 & 1 & 1 & 1 \\ \hline
37 & 0 & 1 & 0 & 0 & 1 & 1 & 1 & 1 & 0 & 1 & 1 & 1 \\ \hline
38 & 1 & 0 & 0 & 1 & 0 & 1 & 1 & 1 & 0 & 1 & 1 & 1 \\ \hline
39 & 0 & 1 & 1 & 1 & 0 & 1 & 1 & 1 & 0 & 1 & 1 & 1 \\ \hline
\end{tabular}
    \caption{The physical configurations in the 6-site $Z_2$ LGT coupled to one flavor of staggered fermions in 1+1 D. The state labeled by `0' is the strong-coupling vacuum, $\ket{\Omega}_0$.}
    \label{tab:physical_z2}
\end{table}
\begin{table}[ht!]
    \centering
    \begin{tabular}{|c||*{12}{c|}}\hline
Physical- & \multirow{3}{*}{$\ket{0}_{f}$} & \multirow{3}{*}{$\ket{0}_{b}$} & \multirow{3}{*}{$\ket{1}_{f}$} & \multirow{3}{*}{$\ket{1}_{b}$} & \multirow{3}{*}{$\ket{2}_{f}$} & \multirow{3}{*}{$\ket{2}_{b}$} & \multirow{3}{*}{$\ket{3}_{f}$} & \multirow{3}{*}{$\ket{3}_{b}$} & \multirow{3}{*}{$\ket{4}_{f}$} & \multirow{3}{*}{$\ket{4}_{b}$} & \multirow{3}{*}{$\ket{5}_{f}$} & \multirow{3}{*}{$\ket{5}_{b}$} \\
configura- & & & & & & & & & & & & \\
tion label & & & & & & & & & & & & \\
\hline
0 & 1 & 0 & 1 & 0 & 1 & $-1$ & 0 & 0 & 0 & 0 & 0 & 1 \\ \hline
1 & 1 & $-1$ & 1 & $-1$ & 0 & $-1$ & 1 & $-1$ & 0 & $-1$ & 0 & 0 \\ \hline
2 & 1 & 0 & 1 & 0 & 0 & 0 & 1 & 0 & 0 & 0 & 0 & 1 \\ \hline
3 & 1 & $-1$ & 0 & 0 & 1 & $-1$ & 1 & $-1$ & 0 & $-1$ & 0 & 0 \\ \hline
4 & 1 & 0 & 0 & 1 & 1 & 0 & 1 & 0 & 0 & 0 & 0 & 1 \\ \hline
5 & 0 & 0 & 1 & 0 & 1 & $-1$ & 1 & $-1$ & 0 & $-1$ & 0 & 0 \\ \hline
6 & 0 & 1 & 1 & 1 & 1 & 0 & 1 & 0 & 0 & 0 & 0 & 1 \\ \hline
7 & 1 & $-1$ & 1 & $-1$ & 0 & $-1$ & 0 & 0 & 1 & $-1$ & 0 & 0 \\ \hline
8 & 1 & 0 & 1 & 0 & 0 & 0 & 0 & 1 & 1 & 0 & 0 & 1 \\ \hline
9 & 1 & $-1$ & 0 & 0 & 1 & $-1$ & 0 & 0 & 1 & $-1$ & 0 & 0 \\ \hline
10 & 1 & 0 & 0 & 1 & 1 & 0 & 0 & 1 & 1 & 0 & 0 & 1 \\ \hline
11 & 0 & 0 & 1 & 0 & 1 & $-1$ & 0 & 0 & 1 & $-1$ & 0 & 0 \\ \hline
12 & 0 & 1 & 1 & 1 & 1 & 0 & 0 & 1 & 1 & 0 & 0 & 1 \\ \hline
13 & 1 & $-1$ & 0 & 0 & 0 & 0 & 1 & 0 & 1 & $-1$ & 0 & 0 \\ \hline
14 & 1 & 0 & 0 & 1 & 0 & 1 & 1 & 1 & 1 & 0 & 0 & 1 \\ \hline
15 & 0 & 0 & 1 & 0 & 0 & 0 & 1 & 0 & 1 & $-1$ & 0 & 0 \\ \hline
16 & 0 & 1 & 1 & 1 & 0 & 1 & 1 & 1 & 1 & 0 & 0 & 1 \\ \hline
17 & 0 & 0 & 0 & 1 & 1 & 0 & 1 & 0 & 1 & $-1$ & 0 & 0 \\ \hline
18 & 1 & $-1$ & 1 & $-1$ & 0 & $-1$ & 0 & 0 & 0 & 0 & 1 & 0 \\ \hline
19 & 1 & 0 & 1 & 0 & 0 & 0 & 0 & 1 & 0 & 1 & 1 & 1 \\ \hline
20 & 1 & $-1$ & 0 & 0 & 1 & $-1$ & 0 & 0 & 0 & 0 & 1 & 0 \\ \hline
21 & 1 & 0 & 0 & 1 & 1 & 0 & 0 & 1 & 0 & 1 & 1 & 1 \\ \hline
22 & 0 & 0 & 1 & 0 & 1 & $-1$ & 0 & 0 & 0 & 0 & 1 & 0 \\ \hline
23 & 0 & 1 & 1 & 1 & 1 & 0 & 0 & 1 & 0 & 1 & 1 & 1 \\ \hline
24 & 1 & $-1$ & 0 & 0 & 0 & 0 & 1 & 0 & 0 & 0 & 1 & 0 \\ \hline
25 & 1 & 0 & 0 & 1 & 0 & 1 & 1 & 1 & 0 & 1 & 1 & 1 \\ \hline
26 & 0 & $-1$ & 1 & $-1$ & 0 & $-1$ & 1 & $-1$ & 0 & $-1$ & 1 & $-1$ \\ \hline
27 & 0 & 0 & 1 & 0 & 0 & 0 & 1 & 0 & 0 & 0 & 1 & 0 \\ \hline
28 & 0 & 1 & 1 & 1 & 0 & 1 & 1 & 1 & 0 & 1 & 1 & 1 \\ \hline
29 & 0 & $-1$ & 0 & 0 & 1 & $-1$ & 1 & $-1$ & 0 & $-1$ & 1 & $-1$ \\ \hline
30 & 0 & 0 & 0 & 1 & 1 & 0 & 1 & 0 & 0 & 0 & 1 & 0 \\ \hline
31 & 1 & $-1$ & 0 & 0 & 0 & 0 & 0 & 1 & 1 & 0 & 1 & 0 \\ \hline
32 & 0 & $-1$ & 1 & $-1$ & 0 & $-1$ & 0 & 0 & 1 & $-1$ & 1 & $-1$ \\ \hline
33 & 0 & 0 & 1 & 0 & 0 & 0 & 0 & 1 & 1 & 0 & 1 & 0 \\ \hline
34 & 0 & $-1$ & 0 & 0 & 1 & $-1$ & 0 & 0 & 1 & $-1$ & 1 & $-1$ \\ \hline
35 & 0 & 0 & 0 & 1 & 1 & 0 & 0 & 1 & 1 & 0 & 1 & 0 \\ \hline
36 & 0 & $-1$ & 0 & 0 & 0 & 0 & 1 & 0 & 1 & $-1$ & 1 & $-1$ \\ \hline
37 & 0 & 0 & 0 & 1 & 0 & 1 & 1 & 1 & 1 & 0 & 1 & 0 \\ \hline
\end{tabular}
    \caption{The physical configurations in the 6-site $U(1)$ LGT coupled to one flavor of staggered fermions in 1+1 D with an electric-field cutoff $\Lambda=1$. The state labeled by `27' is the strong-coupling vacuum, $\ket{\Omega}_0$.}
    \label{tab:physical_u1}
\end{table}

\end{document}